\documentclass[a4paper,english,ngerman,bibliography=totoc]{article}
\usepackage[utf8]{inputenc}

\usepackage{siunitx}
\usepackage{subfigure}
\usepackage{bm}
\usepackage{moreverb}
\usepackage{diagbox}
\usepackage{multirow}
\usepackage{authblk}
\usepackage{amsmath}
\usepackage{pdfpages}
\usepackage{geometry}
\usepackage{babel}

\usepackage[title]{appendix}

\usepackage{algorithm}
\usepackage{algpseudocode}

\usepackage[driverfallback=dvipdfm,colorlinks,bookmarksopen,bookmarksnumbered,citecolor=black,linkcolor=black,urlcolor=black]{hyperref}

\usepackage{tikz}
\usetikzlibrary{
  calc,%
  trees,%
  positioning,%
  arrows,%
  chains,%
  shapes.geometric,%
  decorations.pathreplacing,%
 decorations.pathmorphing,%
  shapes,%
  matrix,%
 shapes.symbols,%
 intersections,%
  fit,%
  patterns,%
  angles,%
  quotes%
 }
\usepackage{pgfplots}
\pgfrealjobname{Feasibility_3D_cargo_vessel}
\usepackage{pgfkeys}
\pgfplotsset{compat=newest}
\pgfkeys{/pgf/number format/1000 sep=}
\usepackage{pgfplotstable}
\pgfplotsset{
/pgfplots/colormap={gray}{rgb255=(0,0,0) rgb255=(255,255,255)}
}
\pgfplotsset{
/pgfplots/colormap={grayinv}{rgb255=(255,255,255) rgb255=(0,0,0)}
}

\tikzstyle{startstop} = [rectangle, rounded corners, minimum width=3cm, minimum height=1cm,text centered, draw=black]
\tikzstyle{io} = [trapezium, trapezium left angle=70, trapezium right angle=110, minimum width=3cm, minimum height=1cm, text centered, draw=black]
\tikzstyle{arrow} = [thick,->,>=stealth]
\tikzstyle{decision} = [diamond, draw,minimum width=2.5cm,minimum height=1.5cm text width=4.5em, text centered, inner sep=0pt]

\definecolor{color1}{HTML}{0000AD} 
\definecolor{color2}{HTML}{FF4500} 
\definecolor{color3}{HTML}{FFA500} 
\definecolor{color4}{HTML}{00BB00} 
\definecolor{color5}{HTML}{9400D3} 
\definecolor{color6}{HTML}{800000} 
\definecolor{color7}{HTML}{000000} 
\definecolor{color8}{HTML}{0000FF} 
\definecolor{color9}{HTML}{FF0000} 
\definecolor{color10}{HTML}{11CCEE} 
\definecolor{color11}{HTML}{606060} 

\tikzset{%
    body/.style={inner sep=0pt,outer sep=0pt,shape=rectangle,draw,thick,pattern=north east lines wide},
    dimen/.style={<->,>=latex,thin,every rectangle node/.style={fill=white,midway}},
    symmetry/.style={dashed,thin},
}

\tikzstyle{line1} = [color=color7,semithick]
\tikzstyle{line2} = [color=color2,densely dotted,thick]
\tikzstyle{line3} = [color=color1,densely dashed,thick]
\tikzstyle{line4} = [color=color4,dash dot,thick]
\tikzstyle{line5} = [color=color5,dash dot dot,thick]
\tikzstyle{line6} = [color=color6,densely dotted,thick]
\tikzstyle{line7} = [color=color10,densely dashed,thick]
\tikzstyle{line8} = [color=color3,thick]

\tikzstyle{line10} = [color=color9,densely dotted,thick]
\tikzstyle{line11} = [color=color12,dash dot,thick]
\tikzstyle{line12} = [color=color13,dash dot,thick]

\tikzstyle{line13} = [color=color4,dash dot,line width=1pt]
\tikzstyle{line14} = [color=color8,densely dashed,line width=1pt]

\tikzstyle{line15} = [color=color1,densely dashed,thick]
\tikzstyle{line16} = [color=color12,dash dot,thick]
\tikzstyle{line17} = [color=color6,densely dotted,thick]

\tikzstyle{mark1} = [color=color7,mark=x,mark size=2pt,mark options=solid,semithick]
\tikzstyle{mark2} = [color=color2,mark=square,mark size=2pt,mark options=solid,semithick]
\tikzstyle{mark3} = [color=color1,mark=triangle,mark size=2pt,mark options=solid,semithick]
\tikzstyle{mark4} = [color=color4,mark=o,mark size=2pt,mark options=solid,semithick]
\tikzstyle{mark5} = [color=color2,mark=square*,mark size=2pt,mark options=solid,semithick]
\tikzstyle{mark6} = [color=color1,mark=triangle*,mark size=2pt,mark options=solid,semithick]
\tikzstyle{mark7} = [color=color4,mark=*,mark size=2pt,mark options=solid,semithick]
\tikzstyle{mark8} = [color=color2,mark=x,mark size=2.5pt,mark options=solid,semithick]
\tikzstyle{mark9} = [color=color1,mark=diamond,mark size=2pt,mark options=solid,semithick]
\tikzstyle{mark10} = [color=color4,mark=asterisk,mark size=2.5pt,mark options=solid,semithick]
\tikzstyle{mark11} = [color=color7,mark=*,mark size=2pt,mark options=solid,semithick]
\tikzstyle{mark12} = [color=color5,mark=*,mark size=2pt,mark options=solid,semithick]
\tikzstyle{mark13} = [color=color7,mark=*,mark size=1pt,mark options=solid,semithick]
\tikzstyle{mark14} = [color=color3,mark=x,mark size=2.5pt,mark options=solid,semithick]

\tikzstyle{mark15} = [color=color7,mark=*,mark size=0.75pt,mark options=solid,semithick]
\tikzstyle{mark16} = [color=color6,mark=*,mark size=1pt,mark options=solid,semithick]
\tikzstyle{mark17} = [color=color8,mark=x,mark size=2.5pt,mark options=solid,semithick]
\tikzstyle{mark18} = [color=color8,mark=x,mark size=4.0pt,mark options=solid,semithick]

\pgfplotsset{major grid style={densely dotted}} 

\newcommand\BibTeX{{\rmfamily B\kern-.05em \textsc{i\kern-.025em b}\kern-.08em
T\kern-.1667em\lower.7ex\hbox{E}\kern-.125emX}}

\begin{document}

\title{Monolithic 3D numerical modeling of granular cargo movement on bulk carriers in waves}
\author[1]{W. D\"usterh\"oft-Wriggers}
\author[1]{T. Rung}
\affil[1]{Hamburg University of Technology, Hamburg, Germany}
\affil[]{corresponding author: wibke.wriggers@tuhh.de}

\date{}

\renewcommand\Affilfont{\itshape\small}
\maketitle


\begin{abstract}
A novel monolithic approach for simulating vessels in waves with granular cargo is presented using a Finite Volume framework. This model integrates a three-phase Volume of Fluid method to represent air, water, and cargo, coupled with a granular material model. The approach incorporates vessel dynamics by assuming rigid-body motion for the vessel's empty hull. A  three degrees of freedom rigid-body motion solver is applied for a 3D case study. The model also includes inviscid far-field boundary conditions facilitating the generation of linear waves approaching the vessel and applies a rigid-perfectly plastic material model for the granular phase. The model's efficacy is demonstrated through a validation of the three-phase Volume of the Fluid method, a verification of the granular material model, and finally, a reconstruction of the "Jian Fu Star" incident in a fully 3D simulation. This integrated approach is a feasibility study for investigating bulk carrier accidents, offering a powerful tool for maritime safety analysis and design optimization.


\end{abstract}


\vspace{-6pt}

\section{Introduction}
\vspace{-2pt}
This paper introduces a fully coupled 3D three-phase model for cargo vessels transporting granular material in waves. The results of two validation and verification cases representing single dynamics of the coupled model and a complete 3D case application study rebuilding the conditions of a vessel sinking incident are presented. The background for the undertaken feasibility study is the accumulated accidents of bulk carriers transporting unsaturated ores between 2009 and 2019, summarized in table \ref{LostVesselsList}, which resulted in many casualties. Further ships transporting nickel or iron ore had to abort their voyage or take stability-enhancing measures to avoid capsizing and one of these incidents is described exemplarily in Lee \cite{Lee2017}. A general description of the problem of carriage of nickel and iron ore on vessels can be found in Rose \cite{Rose2014} and Munro et al. \cite{Munro2016}. Also, historical accidents such as the sinking of the vessel "Melanie Schulte" in 1952 (cf. Teutsch et al. \cite{TeutschGroll}) could be related to cargo instability issues studied in this work.
\\
The effect that was held accountable for the numerous losses of bulk carriers is called liquefaction, which is a sudden viscous behavior of the loaded unsaturated granular material. Geotechnical liquefaction occurs when the forces resulting from pore pressure exceed the inter-particle forces and is mainly described in the context of earthquakes, e.g., Bian et al. \cite{Bian2009}, Di et al. \cite{Di2004, Di2004Paper}, Elgamal et al. \cite{Elgamal2002} and Unno et al. \cite{Unno2008}. The liquefaction of unsaturated granular material is a challenging mechanism to simulate as seen in Wobbes et al. \cite{wobbes2017modeling} and Airey et al. \cite{Airey2021}. Promising research, including the fluidization of granular material by pore gas pressure in the Finite Volume method, can be found in Chupin et al. \cite{chupin2024non}. Current research by Hoang et al. \cite{hoang2024development, hoang2024sph} applying SPH-based methods to solve geomechanical problems could be extended for future applications on the application of bulk cargo transport.\\
Based on the existing data from accidents and incidents, the cargo failure mechanism responsible for the incidents needs to be secured. Besides geotechnical liquefaction, cargo shifting and the so-called "wet base" (cf. TWG \cite{TWG2013},  Rose \cite{Rose2014}) or "dynamic separation" (cf. Ferauge et al. \cite{ferauge2019liquefaction}) mechanisms can lead to stability issues of the vessels.\\
A "wet base" occurs if the water inside the cargo migrates towards the bottom and a fully saturated layer of cargo forms at the bottom of the cargo hold. The highly saturated bottom layer starts moving, induced by the vessel motion, and the more stable drier cargo slides around on the wet cargo layer.\\
The theory of "dynamic separation" is explained in Ferauge et al. \cite{ferauge2019liquefaction} and assumes that a highly saturated cargo layer occurs on top of the cargo pile either by compression of the cargo, which pumps the internal water upwards or by external water coming overboard. Experience reports of captains suggest that water was found on top of the cargo piles and in the corners of the holds on vessels where unstable cargo led to emergency measures.\\
Nevertheless, pure sliding of the iron and nickel ore cargo can also be the reason for the losses of the vessels since the material's cohesion and repose angle depend on the cargo material's saturation level. Most lost vessels are reported to have left from ports in regions where the monsoon rain was drenching the cargo before the vessel left. Sliding of the cargo can also occur in dry bulk cargos, e.g., wood or grains, and the presented method can be applied to incidents related to these cargoes as well, e.g., the incident of the "Yong Feng" in 2021.\\
Therefore, the present approach does not assume geotechnical liquefaction but uses a classical perfectly plastic material model described in \cite{DuesWri2024} depending on the material's cohesion and angle of repose to represent the cargo behavior. Applying material parameters responding to the material information given in the incident reports, this approach can represent the shifting of cargo during the incidents. Saturation-dependent cohesion and angle of repose can be applied to the presented model.
 \begin{table}[t]
\caption{List of bulk carriers carrying nickel or iron ore lost due to cargo shift/ liquefaction from $2009$ until $2024$ as reported in the "Bulk Carrier Casualty Reports" of INTERGARGO \cite{INTERC2019, INTERC2021}.}
\begin{center}
\begin{tabular}{|c|c|c|c|c|c|c|}
\hline
Vessel name&$DWT$ [t]&$LoA$ [m]&$B$ [m]&year&cargo&loss of lives\\
\hline
\hline
Black Rose&37657&187.7&28.4&2009&Indian iron ore fines&1\\
\hline
Asian Forest&14434&128.0&20.0&2009&Indian iron ore fines&0\\
\hline
Jian Fu Star&45108&189.8&31.3&2010&Indonesian nickel ore&13\\
\hline
Nasco Diamond&56893&185.6&32.3&2010&Indonesian nickel ore&22\\
\hline
Hong Wei&50149&189.8&32.3&2010&Indonesian nickel ore&10\\
\hline
Vinalines Queen&56040&190.0&32.3&2011&Indonesian nickel ore&22\\
\hline
Harita Bauxite&48891&192.0&32.0&2013&Indonesian nickel ore&15\\
\hline
Trans Summer&56824&190.0&32.0&2013&Indonesian nickel ore&0\\
\hline
Bulk Jupiter&56009&190.0&32.3&2015&Malaysian Bauxite&18\\
\hline
Emerald Star&57367&190.0&32.0&2017&Indonesian nickel ore&10\\
\hline
Nur Allya&52378&190.0&32.0&2019&Indonesian nickel ore&27\\
\hline
\end{tabular}
\end{center}
\label{LostVesselsList}
\end{table}\\
Other models have been developed to study the cargo behavior on bulk carriers, some of which are similar to the approach of the present model. In Zou et al. \cite{Zou2013}, a level-set-based Finite Volume (FV) free-surface method is used to study the sloshing of highly viscous fluids in rectangular tanks, missing a material model representing the granular phase. A Discrete Element Method (DEM) based model for simulating liquefaction on vessels has been developed by Ju et al. \cite{Ju2018,Ju2019}. This innovative approach provides a granular-level analysis of cargo behavior under maritime conditions. In a related study, Zhang et al. \cite{Zhang2019} employed a coupled non-Newtonian fluid model with a simplified body surface method to investigate vessel responses. Their research examined the intricate relationships between wave frequencies, amplitudes, and resulting ship motions. Wang et al. \cite{Wang2022} applied the DEM methods to explore the influence of material parameters on cargo movement. This study elucidated the critical role of specific cargo properties in determining the risk and extent of potential shifts during transit. Complementing these efforts, Wu et al. \cite{Wu2022} conducted a comprehensive analysis of liquefaction risk utilizing advanced 3D vessel simulations. These collective studies demonstrate the growing sophistication of computational models in addressing the challenges of maritime cargo transport. In the present work, these studies will be expanded by a fully 3D model of a vessel with granular cargo in wave conditions derived from an incident report.\\
 In this paper, the mathematical model and numerical method are first described. Validation cases of the three-phase and granular material models are presented, and the results of a 3D case study of the coupled problem are reported. As a 3D case study, the "Jian Fu Star" incident on 27th October 2010 is chosen due to its representative vessel dimensions deduced from table \ref{LostVesselsList} and the incident conditions are taken from the incident report of the "Panama Maritime Authority" \cite{JianFuStar}.
 
\vspace{-6pt}

\section{Mathematical Model}
\label{ThBackgr}
\vspace{-2pt}
The present approach solves the complex multi-physics problem of loaded granular cargo on a vessel in waves monolithically. In addition to solving the momentum and continuity equations for a three-phase flow, a displacement equation is introduced to depict the cargo shift. The hull is modeled as a rigid body that moves freely in three degrees of freedom in waves generated by inviscid far-field boundary conditions and the cargo is an incompressible perfectly plastic granular material.

\subsection*{Three-phase method}
An extension of the two-phase VoF approach, cf. Hirt et al. \cite{Hirt1981} to three-phase flows is introduced and validated in order to be able to represent granular cargo in a vessel in waves. To represent the three immiscible phases of air, granular material, and water with respective volumes $V^A$, $V^G$, and $V^W$, the air mixture fraction $c_A=V^A$/$V$ and soil mixture fraction $c_S=V^G$/$V$ are introduced where $V$ denotes the overall volume $V=V^A+V^G+V^W$. Therefore, the mixture fraction of water is defined by
\begin{equation}
\frac{V^W}{V}=1-c_A-c_S\;.
\label{threephasevf}
\end{equation}
To obtain the mixture fraction of water, a second mixture equation is introduced in the standard FV, VoF pressure correction algorithm presented in Yakubov et al. \cite{Yakubov2015}, V\"olkner et al. \cite{Svenja2017} and D\"usterh\"oft-Wriggers et al. \cite{DuesWri2022}. The two mixture fraction equations will be called air mixture fraction and soil mixture fraction hereafter, but the three-phase VoF method can also be used with arbitrary materials. Assuming the three phases are incompressible and immiscible, the condition $D c_S$/$Dt$$=0$ has to be fulfilled, leading to the soil mixture fraction equation over the complete volume $V$ 
\begin{equation}
\int_{V}\left(\frac{\partial c_S}{\partial t}+\frac{\partial \left( c_S v_i \right)}{\partial x_i}\right) dV=0\;,
\end{equation}
where $v_i$ is the velocity field, including the velocities of all three immiscible phases. Accordingly, the air mixture fraction equation gets 
\begin{equation}
\int_{V}\left(\frac{\partial c_A}{\partial t}+\frac{\partial \left( c_A v_i \right)}{\partial x_i}\right) dV=0\;.
\label{airmix}
\end{equation}
The density $\rho$ and viscosity $\mu$ of the immiscible three-phase flow problem is given by 
\begin{equation}
\rho=c_A \rho^A+ c_S \rho^S + (1-c_A - c_S) \rho^W\;, \\  \\  \\ 
\mu=c_A \mu^A+ c_S \mu^S + (1-c_A-c_S) \mu^W\;,
\label{stoffgesetzDens}
\end{equation}
using the bulk properties denoted $^A$ for the air phase, $^S$ for the soil phase and $^W$ for the water phase.\\
A nonlinear interpolation method based on an arctangent function given in D\"uesterh\"oft-Wriggers et al. \cite{DuesWri2024} is used for the material interpolation for non-cohesive granular materials.

\subsection*{Material model for granular phase}
\label{MatModel}
As described and validated in D\"usterh\"oft-Wriggers et al. \cite{DuesWri2024}, the material model for the granular phase is based on the Drucker-Prager yield criterion for perfectly plastic materials. A variable soil viscosity $\mu^S$ 
\begin{equation}
\mu^S=\mu^S_{min}+\frac{3 \alpha_{\phi} p+k_c}{ \sqrt{2 \dot{\epsilon}_{ij} \dot{\epsilon}_{ij}}}\left(1-e^{\left(-m { \sqrt{2 \dot{\epsilon}_{ij} \dot{\epsilon}_{ij}}}\right)}\right)
\label{VarMuLarese}
\end{equation}
follows from a regularized Bingham fluid model with a minimum viscosity $\mu^S_{min}=10^{-3}$\;Pa\;s and the constant $m$, determining the maximum viscosity. Here $\dot{\epsilon}_{ij}$ is the (traceless) strain rate tensor. The material constants $k_c$ and $\alpha_{\phi}$ are chosen as in Bui et al. \cite{Bui2008}
\begin{equation}
\alpha_{\phi}=\frac{tan \phi}{\sqrt{9+12 tan^2 \phi}}
\end{equation}
\begin{equation}
k_c=\frac{3 C}{\sqrt{9+12 tan^2 \phi}}
\end{equation}
where $C$ denotes the cohesion and $\phi$ the angle of repose of the granular material. 

\subsection*{Momentum equations and continuity equations}
Using the immiscible conditions ($D c_A$/$Dt$$=0$, $D c_S$/$Dt$$=0$) as well as the incompressibility condition for the bulk densities $\rho^A$, $\rho^S$, $\rho^W$, the continuity equation in differential form can be simplified for the incompressible, immiscible two-phase materials problem to the usual zero-divergence condition
\begin{equation}
\int_{V}\left( \frac{\partial v_i }{\partial x_i}\right) dV=0\;.
\label{conti}
\end{equation}
The momentum equations for the three-phase problem can be formulated as 
\begin{equation}
\int_{V} \frac{\partial}{\partial t} \left(\rho v_i\right) dV+\oint_{A} \left(\rho v_i v_j \right) dA_j=\\
- \oint_{A} p dA_i+\int_{V} \rho g_i dV+ \oint_{A} \mu \left(\frac{\partial v_i}{\partial x_j}+\frac{\partial  v_j}{\partial x_i}\right) dA_j 
\label{momentum}
\end{equation}
where $\mu$ includes the material model of the granular phase via Eqn. \ref{stoffgesetzDens}. 

\subsection*{Displacement equation}
An Eulerian displacement equation is added to the equation system to represent the cargo's displacement. The Eulerian displacement $u_i$ is calculated from the material derivative 
\begin{equation}
  \frac{D u_i}{D t}=v^S_i
\end{equation}
of the velocity of the granular material $v^S_i$ which is in this work obtained by multiplication with the soil mixture fraction with the velocity field $v^S_{i}=c^S v_i$, leading to the following convective transport equation 
\begin{equation}
\int_{V}\left(  \frac{\partial u_i}{\partial t}+\frac{\partial c^S v_j u_i}{\partial x_j} \right) dV=\int_{V}\left(  c^S v_i\right) dV\;.
  \label{displvs}
\end{equation}

\subsection*{Rigid body motion}
A rigid body motion is assumed for the vessel's hull, neglecting possible structural deformations. Therefore, a six degrees of freedom motion solver based on quaternions is coupled with the conservation equations. For the translational motion, Newton's second law 
\begin{equation}
F_i=m a_i
\label{Fma}
\end{equation}
is solved in an absolute coordinate system, where $F_i$ is an external force on the center of gravity of the rigid body, $a_i$ is the acceleration and $m$ is the mass of the rigid body. The external force stems from the viscous fluid and is calculated by integrating the pressure and viscous shear stresses over the rigid body's boundary. To obtain the velocity $\dot{r}_i$ and the position $r_i$ of the rigid body's center of gravity, Eqn. \eqref{Fma} is integrated over time.\\
To solve the rotational motion of the rigid body, the angular momentum equation is noted in local, body-fixed coordinates, leading to Euler's gyroscopic equations
\begin{equation}
M_1=I_{11}\;\dot{\omega}_1+ \left(I_{33}-I_{22}\right)  \omega_2\;  \omega_3 
\label{Euler1}
\end{equation}
\begin{equation}
M_2=I_{22}\;\dot{\omega}_2+ \left(I_{11} -I_{33}\right) \omega_3\; \omega_1
\label{Euler2}
\end{equation}
\begin{equation}
M_3=I_{33}\;\dot{\omega}_3+ \left(I_{22} -I_{11}\right) \omega_1\; \omega_2\;.
\label{Euler3}
\end{equation}
Here $M_j$ are the moments acting from the fluid on the rigid body, obtained by integration over the rigid body's boundary, $I_{ij}$ is the local inertia tensor which is constant, and $\omega_i$ is the local angular velocity and $\dot{\omega}_i$ is the local angular acceleration. A quaternion $p_i$ including the four Euler parameters $e_0$, $e_1$, $e_2$ and $e_3$ is used to describe the rotational position, and a second quaternion $\dot{p}_i$ is introduced to describe the rotational velocity. From Euler's gyroscopic equations \eqref{Euler1}, \eqref{Euler2} and \eqref{Euler3}, the angular acceleration $\dot{\omega}_i$ is obtained and then integrated to the local angular velocity $\omega_i$. The rotational velocity quaternion $\dot{p}_i$ is then obtained by relation
\begin{equation}
\dot{p}_i=\frac{1}{2} L^T_{ij} \omega_i\;,
\end{equation}
where the matrix $L_{ij}$ includes the Euler parameters
\begin{equation}
L_{ij}=\begin{bmatrix}
   - e_1 & e_0 &  e_3 & - e_2 \\
   - e_2 & - e_3 & e_0 & e_1 \\
   - e_3 & e_2 & - e_1 & e_0\;.
\end{bmatrix}
\end{equation}
Afterward, $\dot{p}_i$ is integrated into $p_i$. More details on the implementation of the quaternion motion solver are described in Luo et al. \cite{Luo2012} and Theilen \cite{Theilen2020} and a validation study of the six degrees of freedom solver is shown in Ulrich \cite{Ulrich2013}.

\subsection*{Wave Generating Boundary Conditions}
A similar approach to SWENSE (Spectral Wave Explicit Navier Stokes Equation) as in Gentaz et al. \cite{gentaz2004numerical} generates waves by imposing inviscid far-field boundary conditions. The present approach is described in W\"ockner et al. \cite{Woeckner2010} and W\"ockner-Kluwe \cite{Woeckner2013} and is based on a manipulation of the momentum equations \eqref{momentum} and the air mixture fraction equation \eqref{airmix} in an impingement zone at the boundary of the domain.\\
Instead of the usual formulation of the linear equation system
\begin{equation}
 A^{\Theta}_P\; \Theta_P + \sum_{NB} A^{\Theta}_{NB}\; \Theta_{NB} = S^{\Theta}_P
\end{equation}
where $\Theta_P$ represents a general variable in the cell centre $(\cdot)_P$, $S^{\Theta}_P$ denotes the source term, $A^{\Theta}_P$ marks the main diagonal coefficient and $\sum_{NB} A^{\Theta}_{NB}\; \Theta_{NB}$ denotes the implicit contributions from the neighbouring cells $NB$, following formulation
\begin{equation}
\Bigl \lbrack 1+\beta_w \; \alpha_w(x_w)\;\Bigr \rbrack\; A^{\Theta}_P\; \Theta_P + \sum_{NB} A^{\Theta}_{NB}\; \Theta_{NB} = S^{\Theta}_P+ \Bigl \lbrack A^{\Theta}_P\; \beta_w\; \alpha_w(x_w)\;\Theta_P^{*}\Bigr \rbrack
\label{WaveRB}
\end{equation}
is applied for the momentum and air mixture fraction equations. Here, the factor $\beta_w$ of order $10^{-3}$ and the function $\alpha_w$ depending on the distance to the nearest boundary $x_w$ impose the inviscid wave theory variable value $\Theta_P^{*}$ onto the viscous-flow solution. The function $\alpha_w$ applied in the presented case study is defined by
\begin{equation}
\alpha_w(x_w)=\left(1-\frac{x_w}{d}\right)^3
\end{equation} 
where $d$ is the width of the impingement zone. \\
Different wave theories, e.g., JONSWAP or Stokes waves, can be chosen to derive the velocity and air volume fraction solution from the wave potential. In the present work, the Airy wave theory is applied to represent the regular waves chosen to recreate the incident conditions. Therefore, the wave potential in deep water gets
\begin{equation}
\Phi = Re(-i\; c\; \hat{\zeta} e^{-kz}e^{\omega_w t -k x})
\end{equation}
with $c$ being the phase velocity of the wave, $\hat{\zeta}$ the waves amplitude, $k$ the wave number and $\omega_w$ the waves circular frequency. The coordinate system describing the wave is $x$ in the direction the wave is traveling and $z$ in the direction of the wave deflection. The velocity components $v_x$ and $v_z$ are obtained by differentiating the wave potential in $x$ and $z$, respectively. Therefore they become
\begin{equation}
v_x = \omega_w\hat{\zeta} e^{-kz} cos(\omega_w t -k x)
\end{equation}
\begin{equation}
v_z = -\omega_w\hat{\zeta} e^{-kz} sin(\omega_w t -k x)\;.
\end{equation}
The air volume fraction is set according to the free surface deviation $\zeta$ 
\begin{equation}
\zeta(x,t)=\hat{\zeta} cos(\omega_w t -k x)
\end{equation}
obtained by the dynamic and kinematic free surface boundary conditions.

\section{Numerical method}
\label{NUM}
A pressure-based FV formulation using a cell-centered, co-located variable arrangement on unstructured polyhedral grids, cf. Ferziger \cite{Ferziger} is applied for all equations. Using the 6DoF motion solver, multiple possibilities exist to realize the vessel's motion: overset grids as described in V\"olkner et al. \cite{Svenja2017}, adaptive meshes as in Manzke \cite{Manzke2019}, or a mesh moving with the rigid body. In the presented case study, the last option is chosen to avoid strong deformations of cells during expected large ship motions in the adaptive mesh approach and high computational times due to interpolations in the overset approach. Therefore, the mesh moves with velocity $v^m_{i}$ which coincides with the velocity of the rigid vessel $\dot{r}_i$. The conservation equations must consider the mesh velocity which is achieved by using the relative velocity $(v_i-v^m_{i})$ in all transport terms of the conservation equations given in Sec. \ref{MatModel}. The equation system follows to be
\begin{equation}
\begin{split}
\int_{V} \frac{\partial}{\partial t} \left(\rho v_i\right) dV+\oint_{A} \left(\rho \left(v_i-v^m_i\right) v_j \right) dA_j=\\
- \oint_{A} p dA_i+\int_{V} \rho g_i dV+ \oint_{A} \mu \left(\frac{\partial \left(v_i-v^m_i\right)}{\partial x_j}+\frac{\partial  \left(v_j-v^m_j\right)}{\partial x_i}\right) dA_j 
\label{MomEqSys}
\end{split}
\end{equation}
\begin{equation}
\int_{V}\left( \frac{\partial \left(v_i-v^m_i\right)}{\partial x_i}\right) dV=0
\label{contiEqn}
\end{equation}
\begin{equation}
\int_{V}\left(\frac{\partial c_A}{\partial t}+\frac{\partial \left( c_A \left(v_i-v^m_i\right) \right)}{\partial x_i}\right) dV=0
\label{AirMixEqSys}
\end{equation}
\begin{equation}
\int_{V}\left(\frac{\partial c_S}{\partial t}+\frac{\partial \left( c_S \left(v_i-v^m_i\right) \right)}{\partial x_i}\right) dV=0
\label{SoilMixEqSys}
\end{equation}
\begin{equation}
\int_{V}\left(  \frac{\partial u_i}{\partial t}+\frac{\partial c^S \left(v_j-v^m_j\right) u_i}{\partial x_j} \right) dV=\int_{V}\left(  c^S \left(v_i-v^m_i\right) \right) dV\;.
\label{DisplEqSys}
\end{equation}
To solve the given equation system, a segregated SIMPLE pressure correction procedure is applied, which largely follows Ferziger \cite{Ferziger} with details given in Yakubov et al. \cite{Yakubov2015} and V\"olkner et al. \cite{Svenja2017}. The pressure correction equations and flux, velocity and pressure corrections derived from the continuity equation \eqref{contiEqn} can be found in D\"usterh\"oft-Wriggers et al. \cite{DuesWri2024} in detail.\\
The algorithmic procedure to solve the three-phase flow problem, including a granular phase, free surface waves, and a 6DoF rigid body motion, is described in Alg. \ref{ProcedureFreSCo}. The equations are solved in a segregated manner in a second outer iteration loop until a predefined residual threshold is obtained within each time step.
\begin{algorithm}
\caption{Incompressible three-phase flow including granular phase coupled with rigid-body motion and waves}\label{euclid}
\begin{algorithmic}[lines]
\State{Initialize $v^0_i$, $p^0$, $c_A^0$, $c_S^0$, $u^0_i$, $\rho^0$, $\mu^0$, $a^0_i$, $\dot{r}^0_i$, $r^0_i$, $\dot{\omega}^0_i$, $\omega^0_i$}
\State{$n=0$}
\State{$m=0$}
\While{$n \le$ max. number time steps}\Comment{$n$ denotes the current time step}
\State $n=n+1$
\While{residual $<$ residualThreshold}\Comment{$m$ denotes the number of outer iterations}
\State $m= m+1$
\State update properties \Comment{Eqn. \eqref{VarMuLarese}, Eqn. \eqref{stoffgesetzDens}}
\State calculate fluid forces on rigid body
\State solve equations of motion\Comment{Eqn. \eqref{Fma}, Eqn. \eqref{Euler1}, Eqn. \eqref{Euler2}, Eqn. \eqref{Euler3}}
\State move grid
\State solve momentum equations \Comment{Eqns. \eqref{MomEqSys}, \eqref{WaveRB}}
\State solve first stage of pressure correction equation \Comment{cf. \cite{DuesWri2024}}
\State correct pressures, fluxes and velocities \Comment{cf. \cite{DuesWri2024}}
\State solve second stage of pressure correction equation \Comment{cf. \cite{DuesWri2024}}
\State correct pressures \Comment{cf. \cite{DuesWri2024}}
\State solve air mixture fraction equation \Comment{Eqns. \eqref{AirMixEqSys}, \eqref{WaveRB}}
\State solve soil mixture fraction equation \Comment{Eqn. \eqref{SoilMixEqSys}}
\State solve displacement equations  \Comment{Eqn. \eqref{DisplEqSys}}
\EndWhile\label{euclidendwhile}
\State{$v^n_i$, $p^n$, $c_A^n$, $c_S^n$, $u^n_i$, $\rho^n$, $\mu^n$, $a^n_i$, $\dot{r}^n_i$, $r^n_i$, $\dot{\omega}^n_i$, $\omega^n_i$\\
\; \; $\to$\; $v^{n-1}_i$, $p^{n-1}$, $c_A^{n-1}$, $c_S^{n-1}$, $u^{n-1}_i$, $\rho^{n-1}$, $\mu^{n-1}$, $a^{n-1}_i$, $\dot{r}^{n-1}_i$, $r^{n-1}_i$, $\dot{\omega}^{n-1}_i$, $\omega^{n-1}_i$}
\EndWhile\label{secondwhile}
\State update properties \Comment{Eqn. \eqref{VarMuLarese}, Eqn. \eqref{stoffgesetzDens}}
\State \textbf{return} $v^n_i$, $p^n$, $c_A^n$, $c_S^n$, $u^n_i$, $\rho^n$, $\mu^n$, $a^n_i$, $\dot{r}^n_i$, $r^n_i$, $\dot{\omega}^n_i$, $\omega^n_i$ \Comment{defined at output time steps}
\end{algorithmic}
\label{ProcedureFreSCo}
\end{algorithm}
After solving the soil mixture fraction equation, the soil mixture fraction is clipped with $c_S=1-c_A$ when the sum of the air and soil mixture fractions is greater than one. This relation, therefore, also determines the handling at the bifurcation point. The time integration scheme applied in the 3D application case for the rigid body motion is second-order implicit trapezoidal, and high under-relaxation of the rigid body motion is applied.\\
The integrals in the equation system are approximated through a second-order accurate mid-point integration rule. For temporal discretization, a first-order implicit method is applied in the verification and validation cases, and a second-order implicit time stepping is used in the 3D application case. The fluxes are treated with a flux blending scheme, in which 70\%  of the method leverages the precision of second-order central differencing, and a first-order upwind scheme (UDS) is applied as a baseline formula. The convective term in the mixture equations is discretized with the Quadratic Upstream Interpolation for Convective Kinematics (QUICK) scheme, initially introduced by Leonard \cite{Leonard1979}, and diffusion fluxes are obtained from central differences, which employ a deferred correction approach to account for non-orthogonality and face interpolation-related issues.

\section{Validation and verification studies}
\label{validate}
A validation of the three-phase flow method against experimental data is presented as well as a verification case for the granular material which expands the validation and verification cases considered in D\"usterh\"oft-Wriggers et al. \cite{DuesWri2024}.
\subsection*{Three-phase flow}
\label{ThreePhaseVerficiation}
\begin{figure}[htbp]
\centering
\includegraphics[scale=.2]{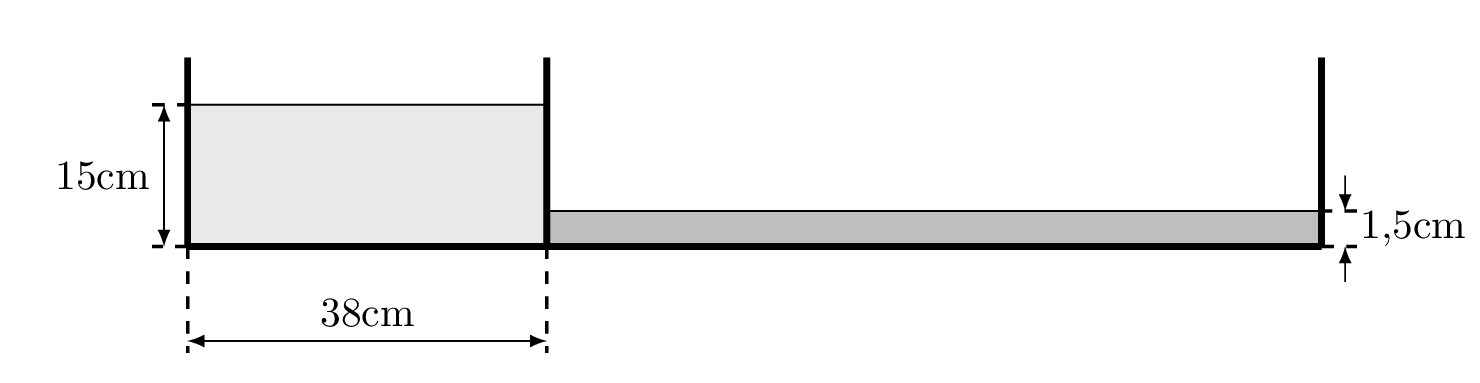}
    \caption{Initial geometry of three-phase dam break experiments by J\'{a}nosi et. al \cite{Janosi2004}}
 \label{initthreephase}
\end{figure}

To validate the three-phase flow model, a dam break experiment by J\'{a}nosi et al. \cite{Janosi2004}, where a higher column of clear water breaks into a layer of colored water, is used as a reference. Besides the clear and colored water, air is considered the third phase in the present model. The dimensions of the initial state, in which the clear water is separated from the colored water by a floodgate, of the experiment are displayed in Fig. \ref{initthreephase}. The rectangular computational domain is chosen to have a height of $h=0.2$\;m and a width of $1.2$\;m with a cell size of $\Delta x_1=\Delta x_2=h$\;/$100$.  At the beginning of the experiment, the floodgate is pulled up, and a time step of $10^{-3}$\;s is used for the simulations.
\begin{figure}[htbp]
\centering
  \subfigure[Experiment J\'{a}nosi et. al \cite{Janosi2004}]{ 
\includegraphics[scale=.23]{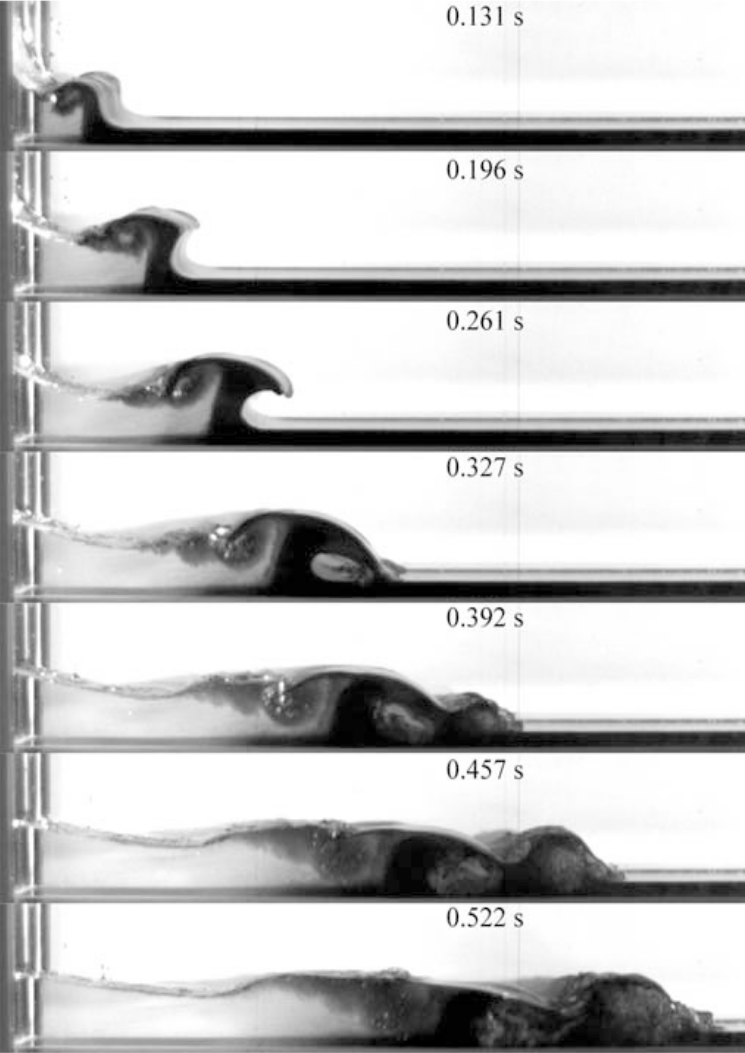}
   }
 \subfigure[Numerical results present method]{ 
 \includegraphics[scale=.098]{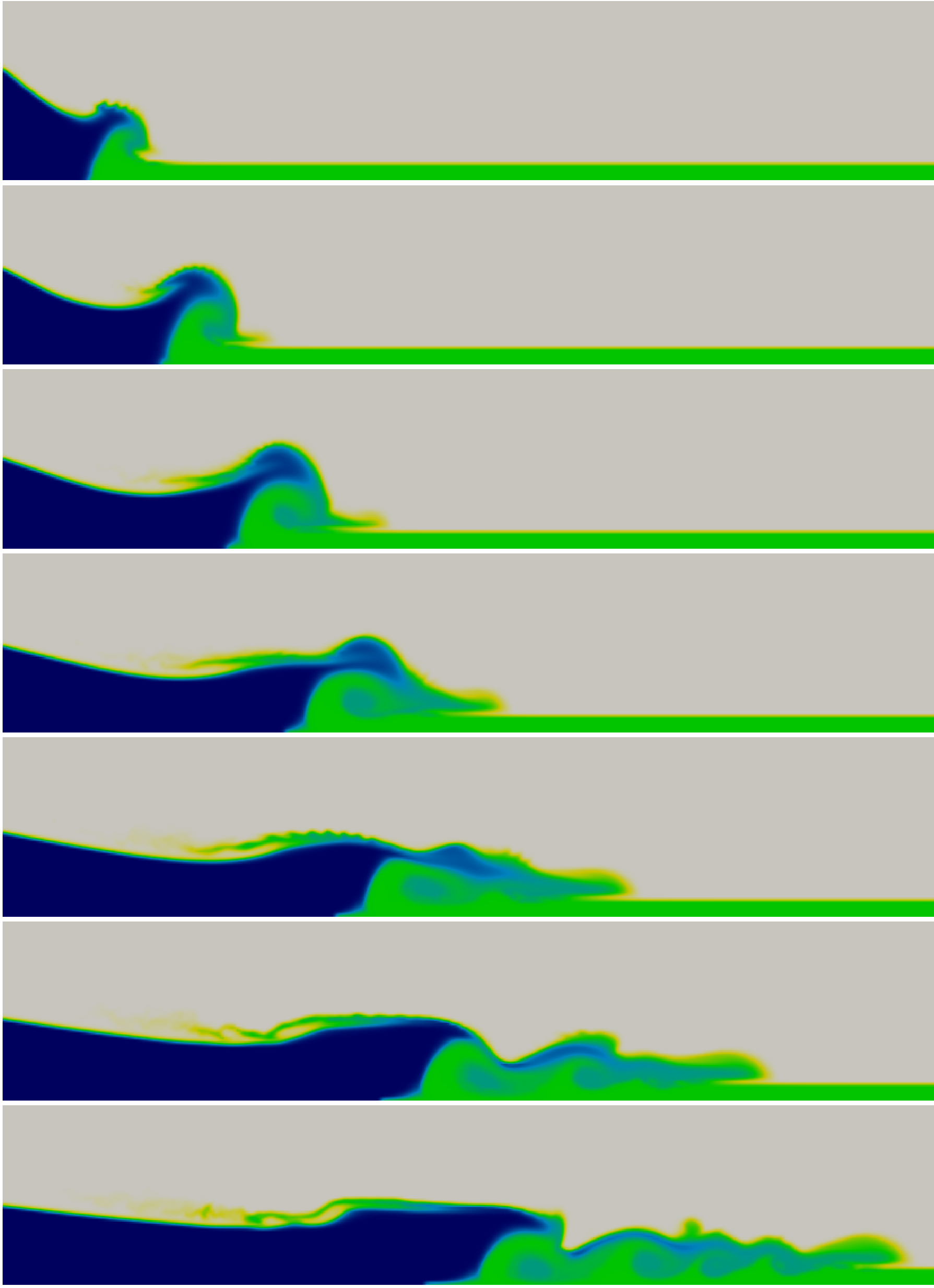}
}
    \caption{Snap shots of experimental results from J\'{a}nosi et al. \cite{Janosi2004} and numerical results from the present three-phase Volume of Fluid method.}
 \label{ThreePhaseResults}
\end{figure}
The results obtained from the present method are visually compared with the experimental data in Fig. \ref{ThreePhaseResults}. In this figure, grey denotes air, green represents the colored water, and blue signifies the clear water.\\
The numerical results demonstrate high concordance with the experimental data regarding the front position across all time steps. However, a notable divergence in breaking behavior is observed at $t=0.131$\;s, attributable to the floodgate opening mechanism in the experimental setup, which is not replicated in the numerical simulations. The velocity profile within the dam exhibits discrepancies when comparing the floodgate opening scenario to the free-breaking dam simulation. Specifically, the numerical results indicate a higher rightward velocity at the dam's apex than the experimental observations. Consequently, a greater proportion of clear water is discernible at the dam's crest in the numerical results, a phenomenon that persists throughout subsequent time steps.\\
Intriguingly, the three-phase VoF method yields a less pronounced representation of the breaking water-free surface morphology at $t=0.196$\;s and $t=0.261$\;s relative to the experimental findings. 

\subsection*{Granular Material}
\begin{figure}[htbp]
  \centering
   \includegraphics[scale=.7]{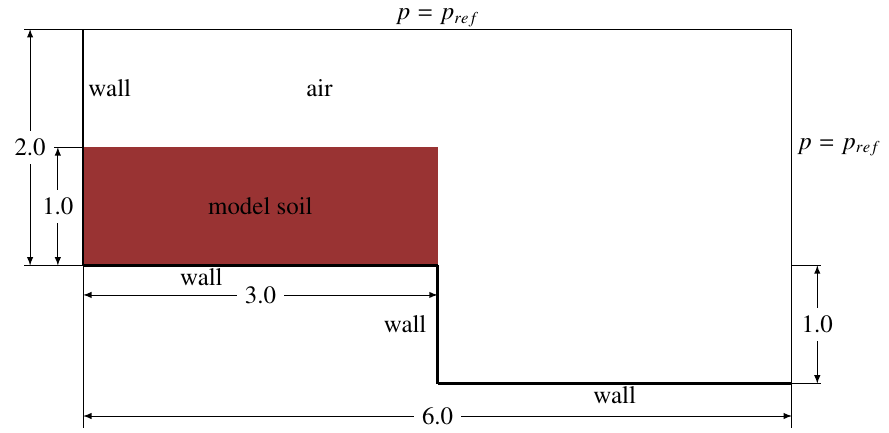}
  \caption{Initial set up of 2D soil dam breaking down a step test case by Larese \cite{Larese2013} with measurements given in meters.}
  \label{LareseAngleOfRepseInit}
\end{figure}

\begin{figure}[htbp]
  \centering
    \subfigure[$t=0.5$\;$\si{\second}$]{ 
    \includegraphics[scale=.7]{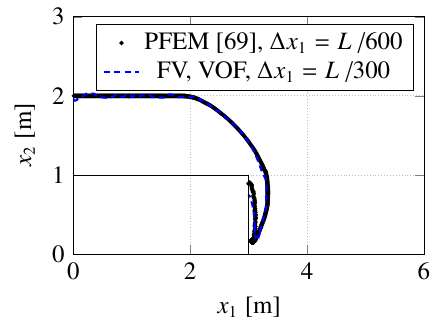}
}
  \subfigure[$t=1.0$\;$\si{\second}$]{ 
    \includegraphics[scale=.7]{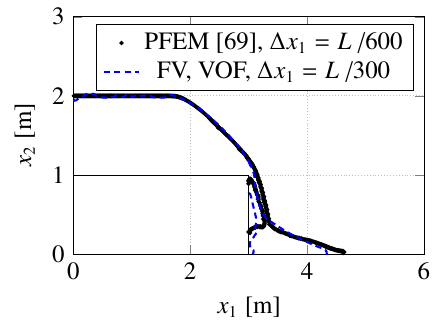}
}
 \subfigure[$t=1.5$\;$\si{\second}$]{ 
    \includegraphics[scale=.7]{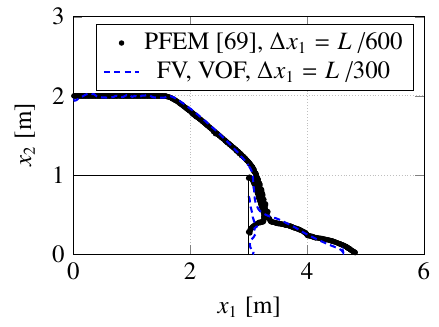}
}
 \subfigure[$t=10.0$\;$\si{\second}$]{ 
    \includegraphics[scale=.7]{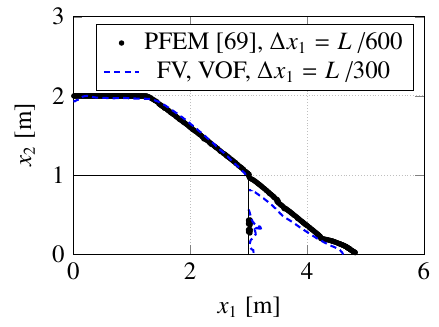}
}

    \caption{Comparison of soil surfaces for a 2D soil dam breaking down a step at four time steps between the present FV, VoF method against pFEM results from Larese \cite{Larese2013}.}
  \label{InternalFriction_Larese_N25}
\end{figure}

A soil dam experiment is conducted to evaluate various angles of repose, following the methodology introduced by Larese \cite{Larese2013}. The experimental setup consists of a soil dam with dimensions of $3$\;m width and $1$\;m height, positioned on a step with a width of $3$\;m. The dam is released at $t=0.0$\;s, initiating the observation period. Fig. \ref{LareseAngleOfRepseInit} illustrates the initial geometry and boundary conditions of the experiment, where the total length $L$ of the computational domain is $6.0$\;m.\\
During the analysis, the following artificial material properties are assigned: soil density $\rho^S=1000$\;kg/m$^3$, minimum soil viscosity $\mu^S_{min}=10^{-6}$\;Pa\;s and reference pressure $p_{ref}=0.0$\;Pa.\\
In this study, results obtained for an angle of repose of $\phi=30^\circ$ are compared with those derived from the particle Finite Element Method (pFEM) by Larese \cite{Larese2013}. For a more comprehensive understanding of the pFEM methodology, readers are directed to the work of O\~{n}ate et al. \cite{Onate2004}. As a yield criterion, following yield function $\Phi$
\begin{equation}
\Phi=\sqrt{J_2}-p\;tan\;\phi -C\;.
\end{equation} including the second invariant of the deviatoric stress $J_2$ is employed to validate results against Larese's findings. Therefore, the primary distinction between the applied granular models lies in their discretization methods and the consideration of single-phase (Larese \cite{Larese2013}) versus two-phase (present study) problems.\\
In Fig. \ref{InternalFriction_Larese_N25}, a comparative analysis of simulation results obtained through two distinct methodologies is presented. The present FV, VoF approach employs a non-linear interpolation of density and viscosity, as extensively discussed by D\"usterh\"oft-Wriggers et al. \cite{DuesWri2024}. This method utilizes a homogenous structured mesh with a uniform grid spacing of $\Delta x_1=\Delta x_2=L$\;/$300$ and a time step of $\Delta t=10^{-3}$\;s. The second approach uses pFEM with an average spatial resolution of $\Delta x_1=\Delta x_2=L$\;/$600$. The key observations are that the granular model discretized with the FV and VoF methods exhibits more vital adherence to the wall during the flow process. Despite the difference in wall adhesion, both methods demonstrate excellent concordance in their overall results. This comparative analysis highlights the robustness of both numerical approaches in simulating granular flow dynamics, while also revealing subtle differences in their treatment of wall interactions.

\section{3D case study of bulk carrier in waves loaded with granular material}
\label{validate}
\begin{figure}[htbp]
\centering
\includegraphics[scale=.5]{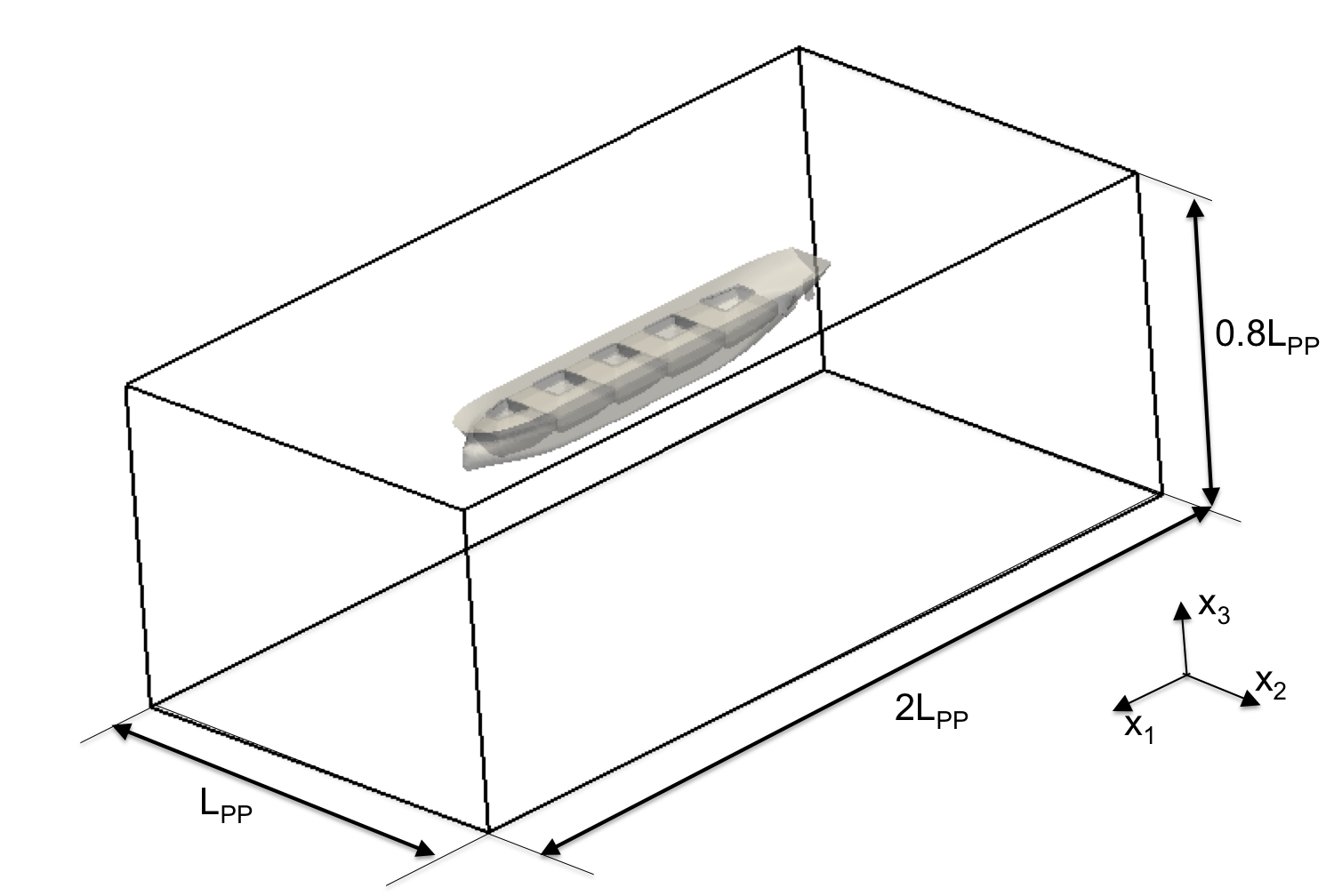}
\caption{Domain dimensions and bulk carrier geometry of 3D cargo vessel in waves.}
\label{DomainMitSchiff}
\end{figure}

\begin{figure}[htbp]
\centering
 \subfigure[Longitudinal section through bulk carrier]{ 
\includegraphics[scale=.37]{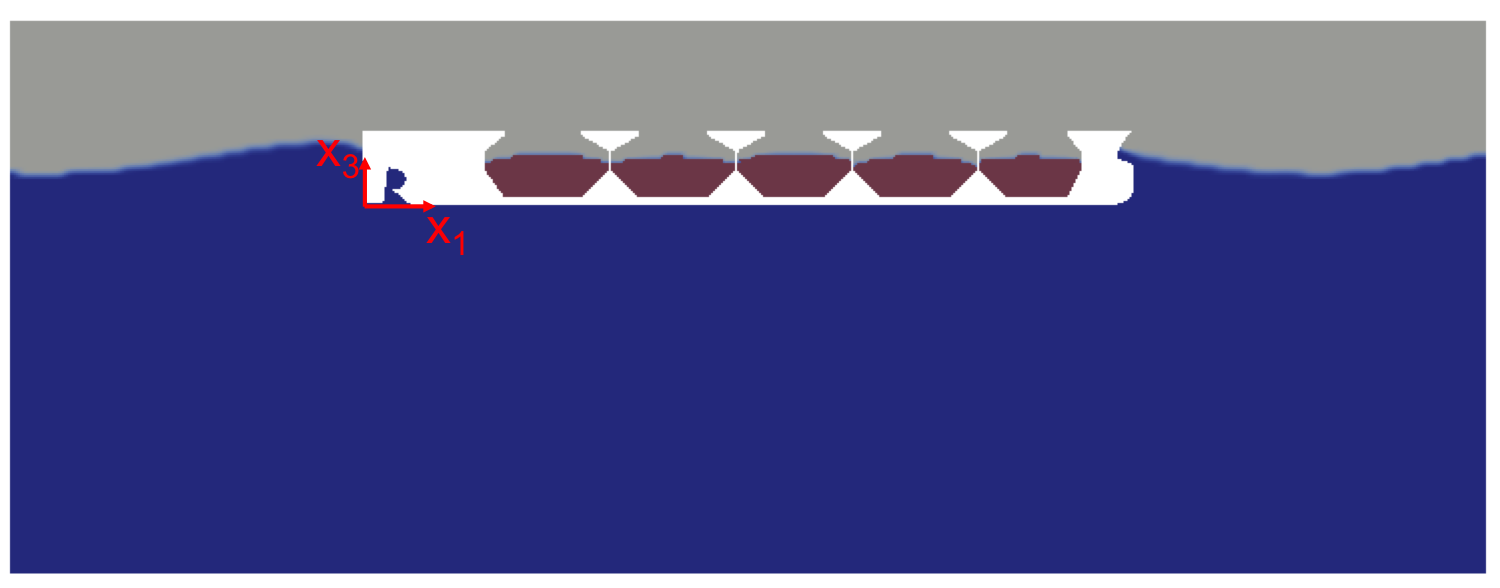}
}
         \subfigure[Cross-section through bulk carrier]{ 
 \includegraphics[scale=.37]{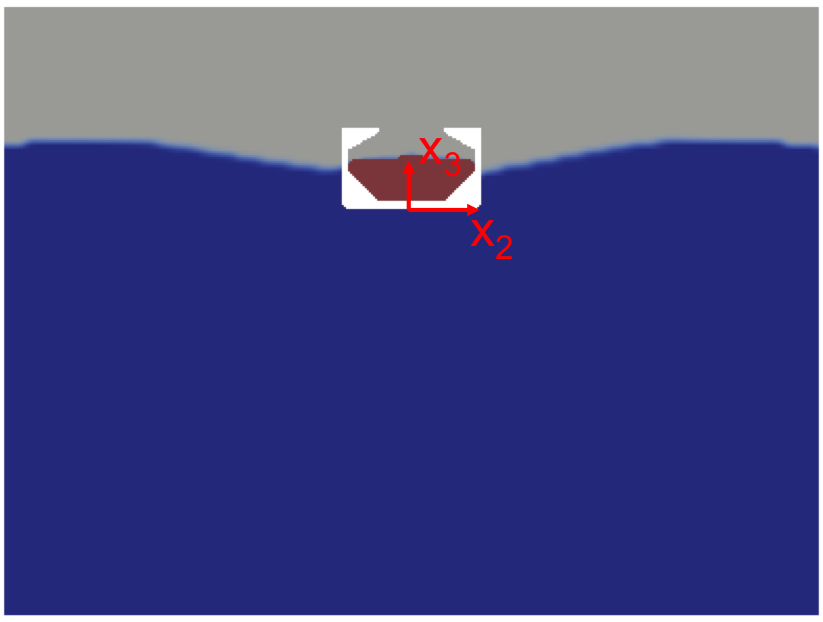}
}
    \caption{3D cargo vessel in waves: Initialization of granular cargo (red), water (blue), and air phases (light grey) is displayed in two different sections through the computational domain.}
 \label{SlicesInitCargoVessel}
\end{figure}

\begin{figure}[htbp]
\centering
  \subfigure[Bulk carrier geometry with initialized cargo]{ 
\includegraphics[scale=.3]{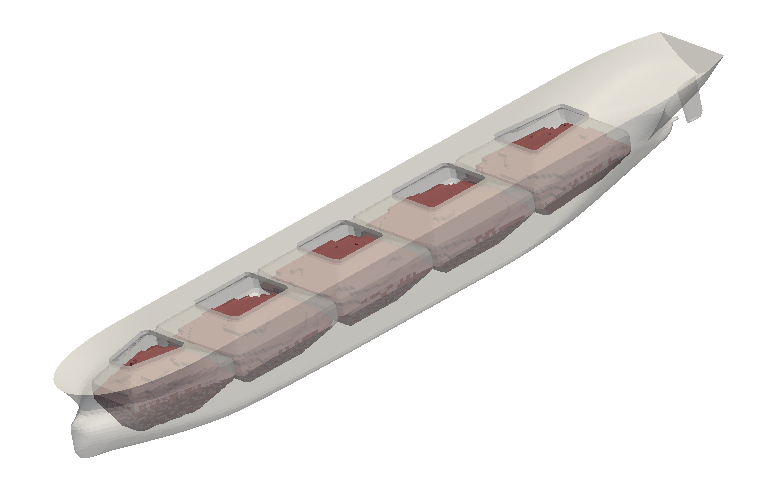}
   }
  \subfigure[Free surface height $h_w$]{ 
\includegraphics[scale=.27]{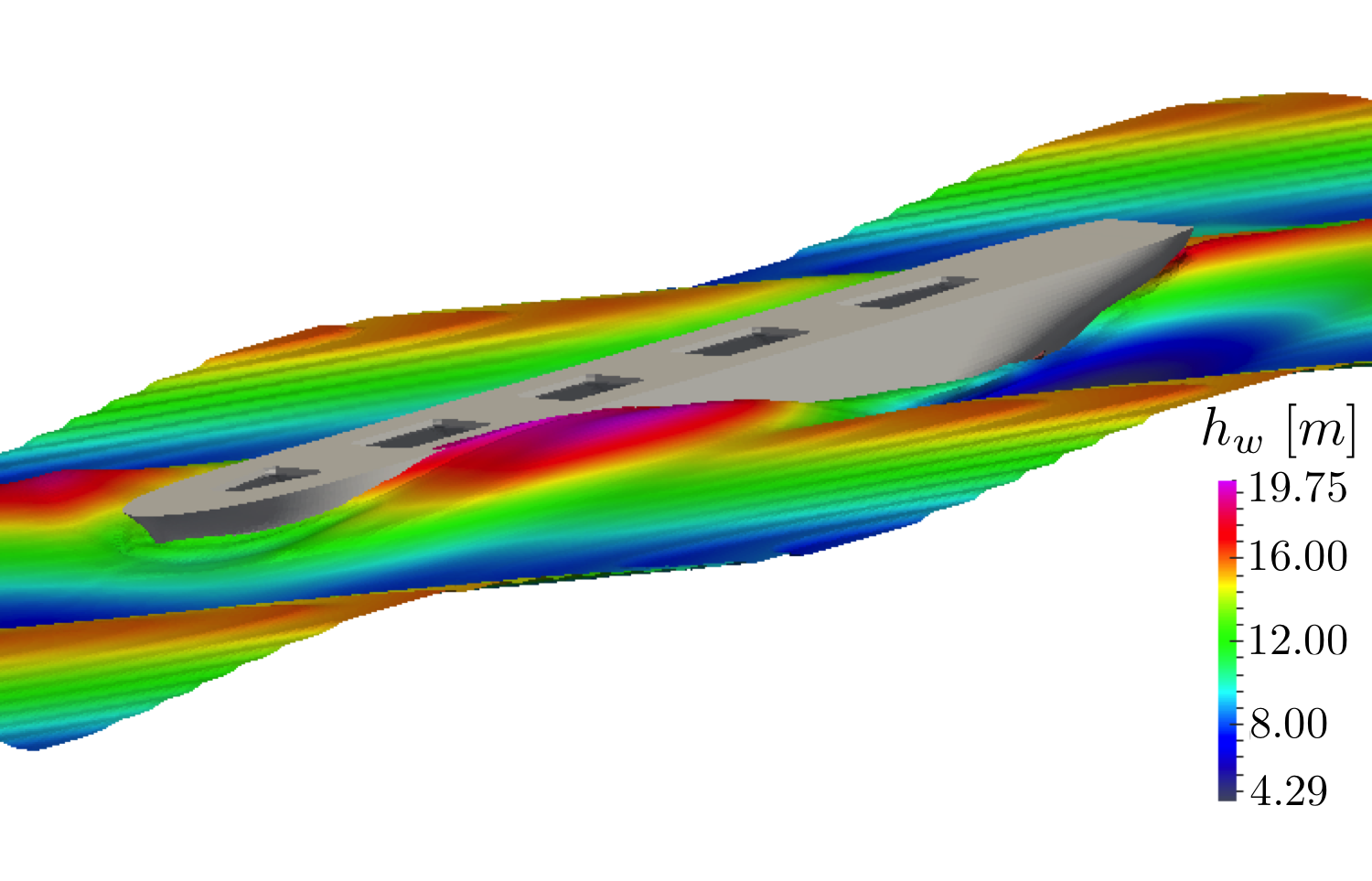}
   }
    \caption{3D cargo vessel in waves: Loaded bulk carrier geometry and wave conditions at an early time step. The free surface height $h_w$ of the water phase refers to the origin of the coordinate system which is positioned at the bottom of the vessel hull.}
 \label{3DInitCargoVessel}
\end{figure}

\begin{figure}[htb]
\centering
\includegraphics[scale=.5]{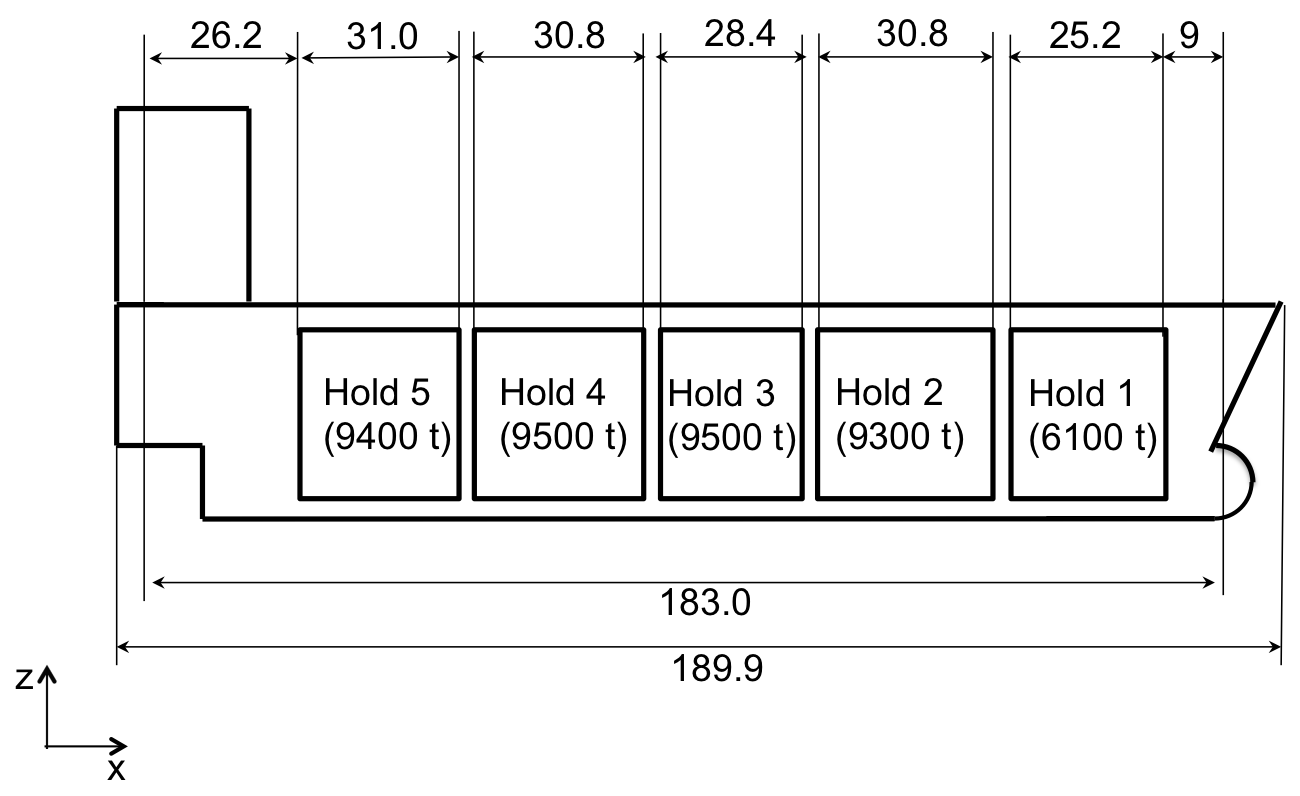}
\caption{Arrangement of holds in bulk carrier "Jian Fu Star". Dimensions are given in meters.}
\label{holdsJianFuStar}
\end{figure}

\begin{figure}[htb]
\centering
\includegraphics[scale=.5]{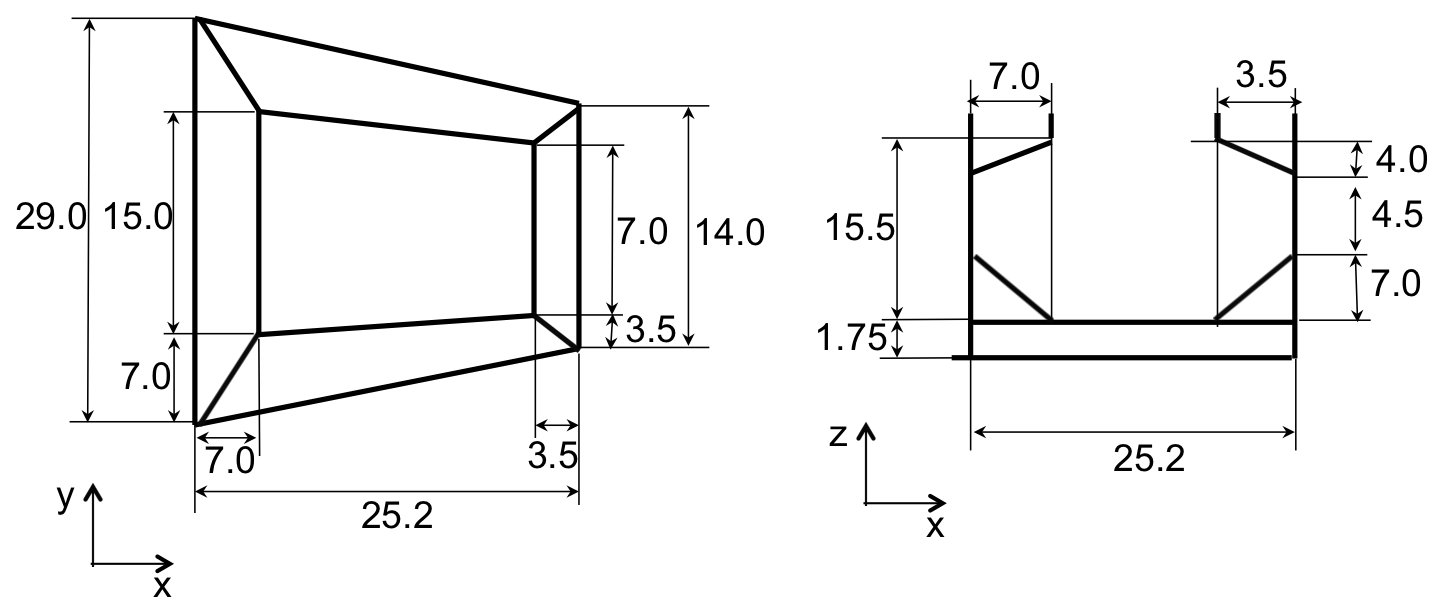}
\caption{Geometry of hold one with dimensions given in metres.}
\label{hold1}
\end{figure}

\begin{figure}[htb]
\centering
\includegraphics[scale=.5]{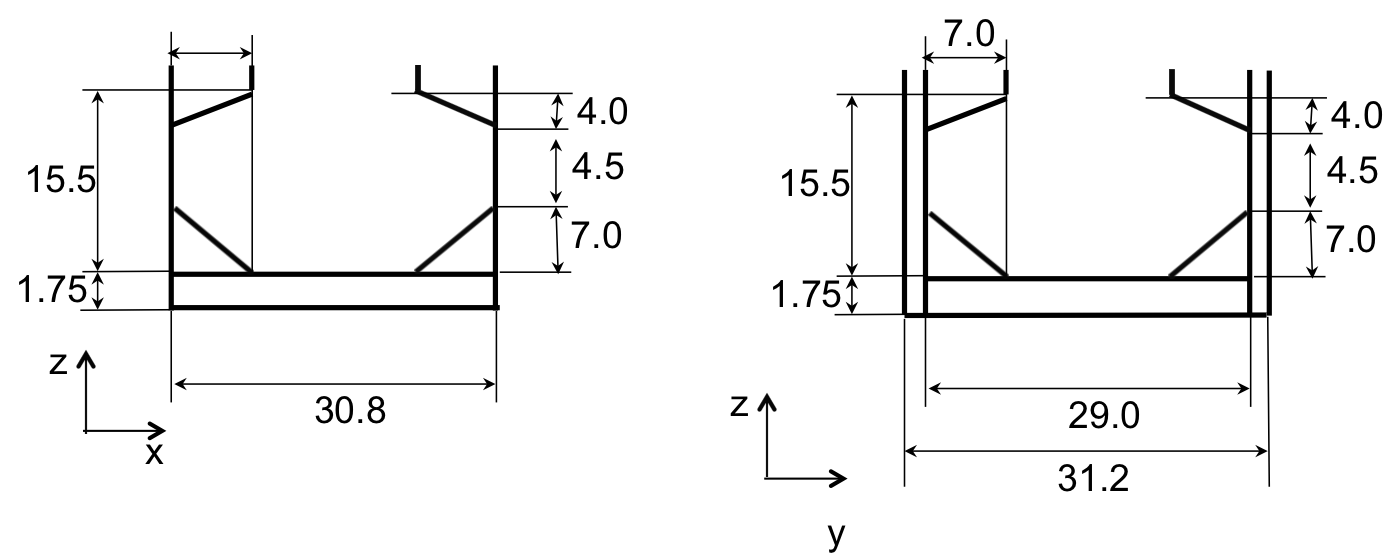}
\caption{Geometry of hold two, hold three and hold four with dimensions given in metres.}
\label{hold243}
\end{figure}

\begin{figure}[htb]
\centering
\includegraphics[scale=.5]{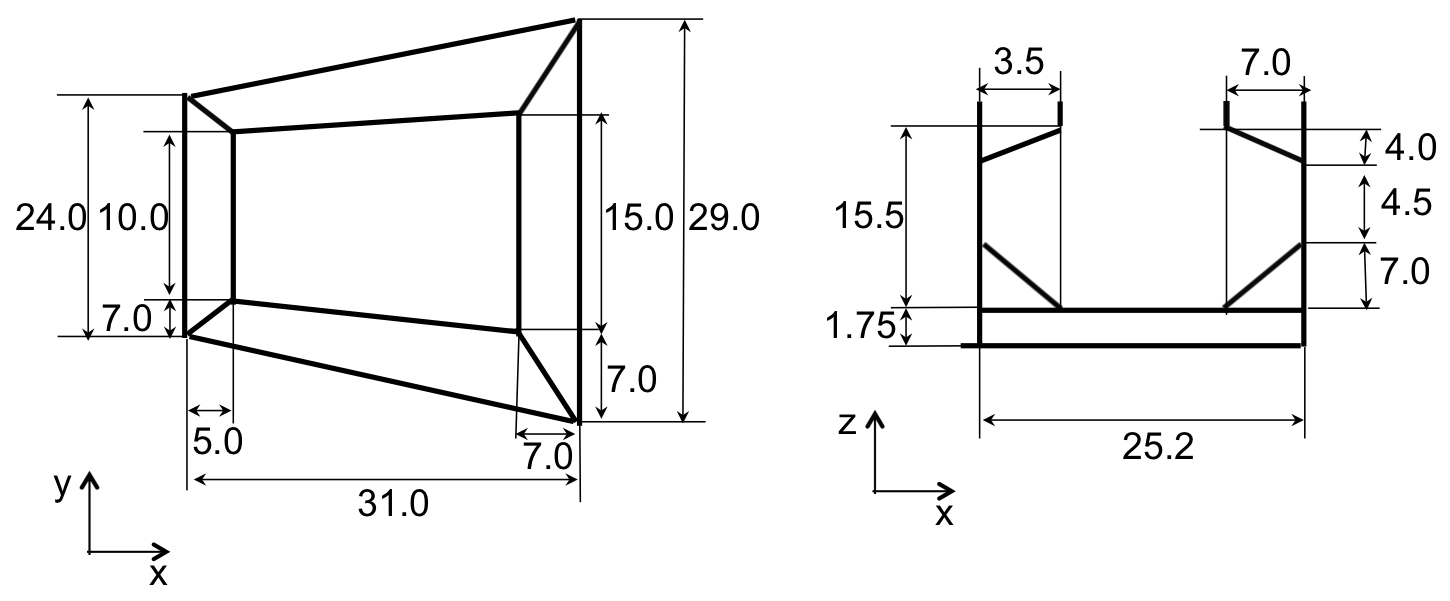}
\caption{Geometry of hold five with dimensions given in metres.}
\label{hold5}
\end{figure}

The sinking of the "Jian Fu Star" occurred on 27th October after 21 hours of high Northerly winds between 5 and 8 Beaufort. It transported Indonesian nickel ore in five cargo holds and had a length over all of $189.8$\;m, a width of $31.2$\;m, and a summer draught of $11.43$\;m. The vessel's rolling motion was moderate, but it was pitching heavily, and water came on the forward deck as described by the survivors in \cite{JianFuStar}. To prevent the surcharge of the engines, the vessel speed was reduced to $4.5$\;kn followed by a sudden list of $5^\circ$ to port at $7$ AM. Ballast water pumping as counteraction was started. However, these measures did not have the desired effect, and the vessel further listed to $10^\circ$. The vessel then rolled around this list angle, taking on high seas on board, and sank $20$ minutes after the first list had occurred. Twelve seamen were rescued, and thirteen seamen were lost.\\
To reconstruct the loading conditions, the soil fraction is initialised according to the stowage plan given in the investigation of the "Panama Maritime Authority" \cite{JianFuStar}, cf.  Fig. \ref{holdsJianFuStar}. A more detailed look at the initial conditions of the three phases, air, water, and granular material, is presented in Fig. \ref{SlicesInitCargoVessel}.\\
Also, the main dimensions of the "Jian Fu Star" are taken from the incident report \cite{JianFuStar} and are partly presented in Fig. \ref{holdsJianFuStar}: length over all $LoA=189.9$\;m, width $B=31.2$\;m and metacentric height $GM=5.0$\;m. A generic bulk carrier hull geometry was scaled to fulfill these properties, resulting in a draft of $D=11.8$\;m, a freeboard of $F=6.7$\;m and a length between perpendiculars $L_{PP}=183.0$\;m. Cargo holds with the dimensions given in Figs. \ref{hold1}, \ref{hold243}, and \ref{hold5} were added to the generic hull geometry corresponding to the arrangement of the cargo compartments as given in Fig. \ref{holdsJianFuStar}. Unlike in reality, the cargo holds are not closed with lids but are connected to the air, creating a contiguous domain for solving the conservation equations. The final ship geometry and the dimensions of the computational domain in relation to the length between perpendiculars are displayed in Fig. \ref{DomainMitSchiff}.\\
As given in the incident report \cite{JianFuStar}, the moisture content $MC$ was certified to be $34.45\%$, and the measured particle size did not exceed $200$\;mm. Since no further test on the materials properties was repeated directly before the "Jian Fu Star" 's departure, the incident report \cite{JianFuStar} concludes that a higher $MC$ during carriage cannot be excluded. In this reconstruction simulation, the cargo properties are selected to match Indonesian nickel ore with a $MC$ of $29\%$ to use a conservative estimation of the natural material properties. From the material property table in the "Guidelines for the Safe Carriage of Nickel Ore" by Class NK \cite{ClassNK} responding material properties for a $MC$ of $29\%$ are obtained, and therefore the angle of repose $\phi$ is set to $4.0^\circ$, the cohesion $C$ is set as $7900.0$\;Pa and the cargo density is set to the wet density of $1700$\;kg/m$^3$. The Indonesian nickel ore is simulated, applying the rigid-perfectly plastic approach with the Drucker-Prager yield criterion given in Sec. \ref{MatModel}. The density of air and sea water are set to $1.185$\;kg/m$^3$ and $1025.0$\;kg/m$^3$ and the viscosities to $1.831\cdot10^{-5}$\;Pa\,s and $1.132\cdot10^{-3}$\;Pa\,s.\\
A three-degree-of-freedom rigid body motion of the vessel is assumed, where the vessel can heave, pitch, and roll. The vessel properties determining the rigid body motion include the position of the center of gravity, the mass, and the moments of inertia of the bulk carrier. In this study, the simulated vessel refers to an empty hull, and the cargo is realized by a continuum whose mass acts on the ship as a pressure force. Therefore, the data for the empty bulk carrier, cf. Tab. \ref{DataEmptyVessel} is derived by subtracting the properties of the loaded cargo from the fully loaded bulk carriers' properties obtained from hydrostatic formulas, the vessel data given by the incident report, and the generic hull model.
\begin{table}[h!]
\caption{Data of empty bulk carrier model "Jian Fu Star" in incident study}
\begin{center}
\begin{tabular}{l l}
\hline
$m$ (mass)&$13634.643$\;t\\
\hline
$CoG$ (center of gravity)&$(92.875|0.0|8.497)$\;m\\
\hline
$I_{11}$ (moment of inertia)& $3.3548749750  \cdot 10^{9}$\;kg m$^2$\\
\hline
$I_{22}$ (moment of inertia)& $4.0278141357 \cdot 10^{10}$\;kg m$^2$\\
\hline
$I_{33}$ (moment of inertia)& $4.5380647175 \cdot 10^{10}$\;kg m$^2$\\
\hline
$I_{23}=I_{32}$ (moment of inertia)& $- 4.3608019573 \cdot 10^{7}$\;kg m$^2$\\
\hline
\end{tabular}
\end{center}
\label{DataEmptyVessel}
\end{table}
As displayed in Fig. \ref{SlicesInitCargoVessel}, the mean level of the free surface between air and water is initialized at the height of $11.8$\;m, following the departure draft $D$. Estimated from the reported weather conditions, a linear Airy wave with a height of $8$\;m and length of $0.57 L_{PP}$ is initialized and taken as a far-field solution in the wave boundary conditions given in Sec. \ref{MatModel}. Since the vessel was sailing to the north-east and the wind direction is reported as northerly, for the incident study the waves are assumed to hit the bulk carrier diagonally from the front with an angle of $48^\circ$ towards the vessel's longitudinal axis. In Fig. \ref{3DInitCargoVessel}, the vessel in waves is depicted for an early time step.\\
For the incident study, a coarse unstructured mesh with $9.07$\;Mio cells used,  resulting in a mean cell size of $\Delta x_1 \approx 2 L_{PP}$\;/$368$. The mesh is evenly distributed due to the expected large rotations of the vessel, where the entire mesh is moved due to rigid body motions. An implicit Crank-Nicholson time discretization is applied to all equations with a time step of $10^{-4}$\;s. To counteract the numerical diffusion induced by the coarse mesh resolution, a non-linear interpolation method described in D\"usterh\"oft-Wriggers et al. \cite{DuesWri2024} is applied to sharpen the material properties at the soil-air interface.\\
The computational domain is characterized by the implementation of high Reynolds wall boundary conditions, which are applied to the vessel hull and all peripheral boundaries, with the exception of the upper boundary. A negative vessel velocity is prescribed as Dirichlet boundary conditions for the velocity vector to simulate the vessel's motion through the fluid. Based on the incident report, a vessel speed of $4.5$\;kn is assumed, resulting in the following dimensionless parameters: Froude number $Fn=0.055$, Reynolds number $Re=3.8\cdot10^{8}$ and Courant number $Co=2.3\cdot 10^{-4}$. These parameters are crucial for characterizing the flow regime and ensuring numerical stability. At the domain's upper boundary, the pressure is set to $0$\;Pa, serving as a reference pressure for the entire system.\\
The simulation incorporates a rigid body motion model, which necessitates a high under-relaxation factor of $0.025$. This adjustment is essential due to the significant pressure and viscous forces exerted on the holds by the rigid component of the plastic material approach. Notably, in the absence of an elastic formulation, the high viscosities required to achieve rigid behavior (with the consistency factor $m_c$ set to $100$) lead to rapidly fluctuating viscous forces at the wall boundaries. This phenomenon underscores the importance of incorporating elasticity in the granular model to more accurately represent the material behavior and improve numerical stability.

\subsection{Results of 3D Feasibility Study with Rigid Perfectly-Plastic Cargo}
\begin{figure}[htbp]
\centering
\includegraphics[scale=.65]{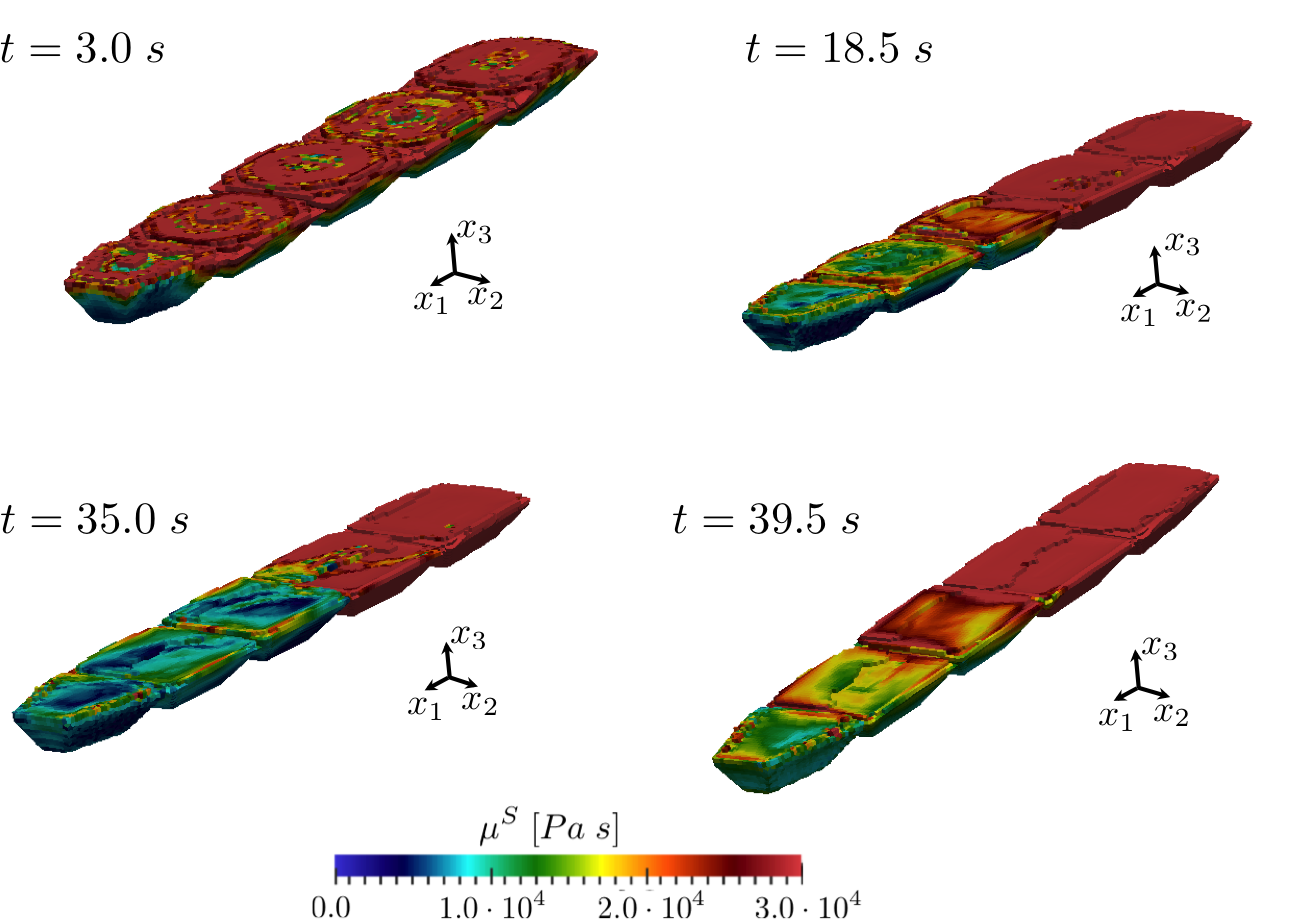}
    \caption{Snap shots of the soil viscosity of the rigid perfectly-plastic cargo at four times during voyage in waves.}
 \label{Holds3DViscos}
\end{figure}

\begin{figure}[htbp]
  \centering
   \includegraphics{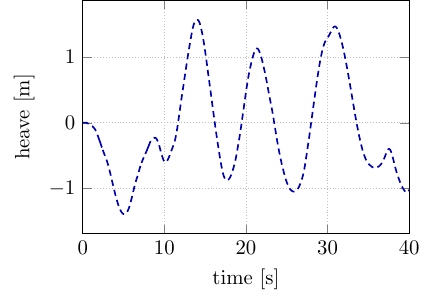}
 \includegraphics{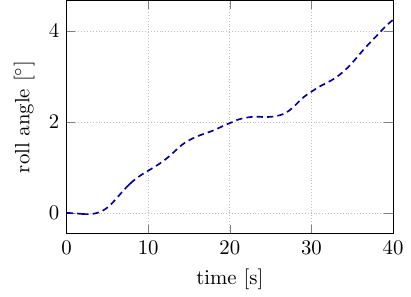}
  \includegraphics{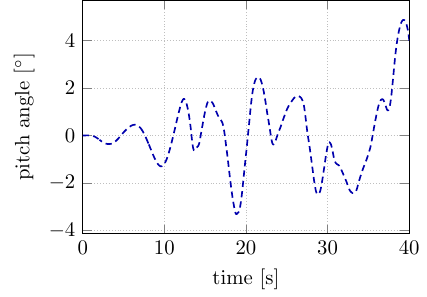}
  \caption{Cargo vessel motions of 3DoF vessel in waves with rigid perfectly-plastic cargo over time.}
  \label{3DMotion}
\end{figure}

\begin{figure}[htbp]
\centering
  \subfigure[$t=6.5$\;s]{ 
  \includegraphics[scale=.18]{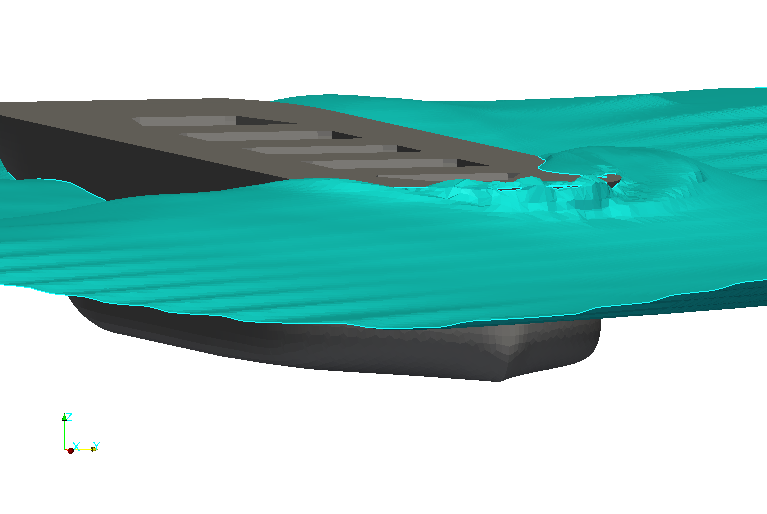}
    }
      \subfigure[$t=8.5$\;s]{ 
  \includegraphics[scale=.18]{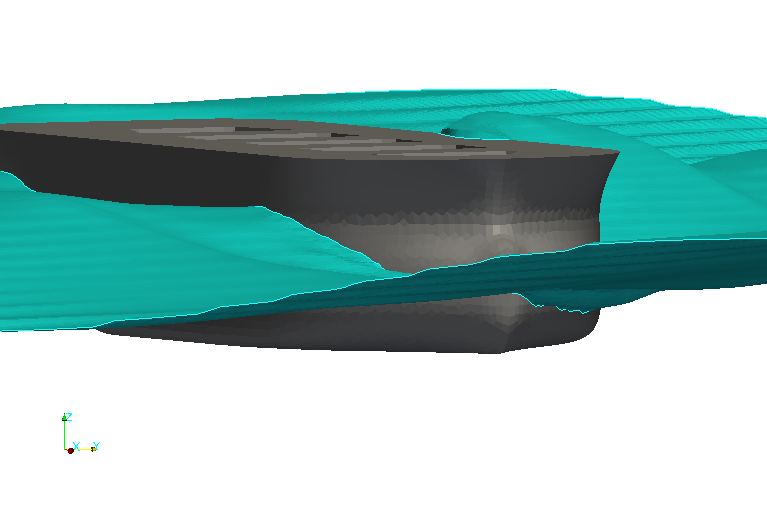}
    }
      \subfigure[$t=20.0$\;s]{ 
  \includegraphics[scale=.18]{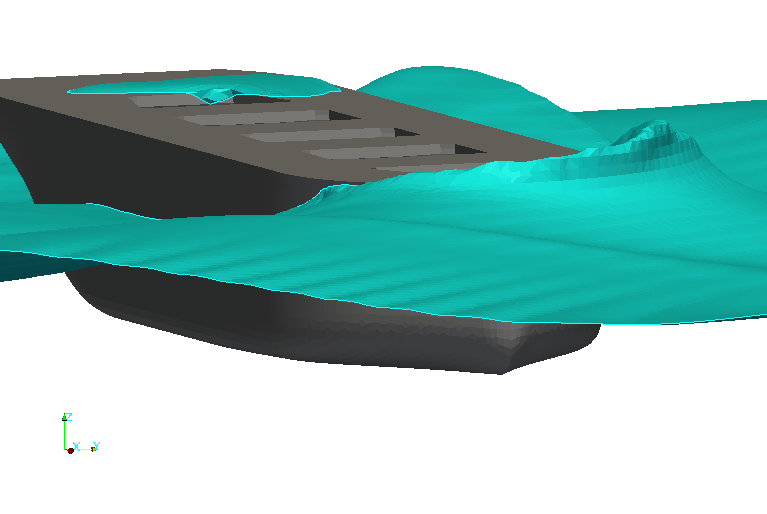}
    }
      \subfigure[$t=25.0$\;s]{ 
  \includegraphics[scale=.18]{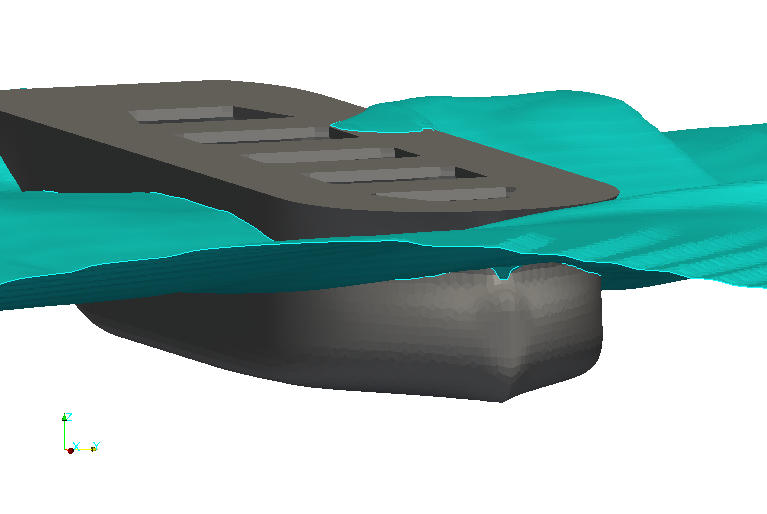}
    }
     \subfigure[$t=37.0$\;s]{ 
  \includegraphics[scale=.18]{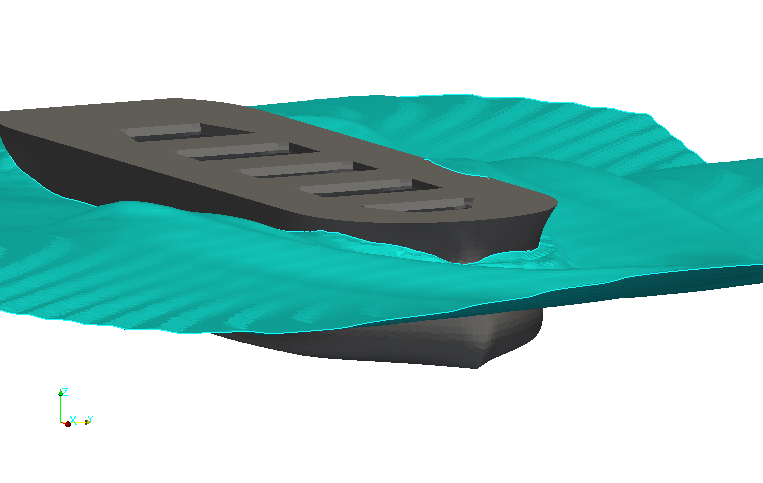}
    }
         \subfigure[$t=39.5$\;s]{ 
  \includegraphics[scale=.18]{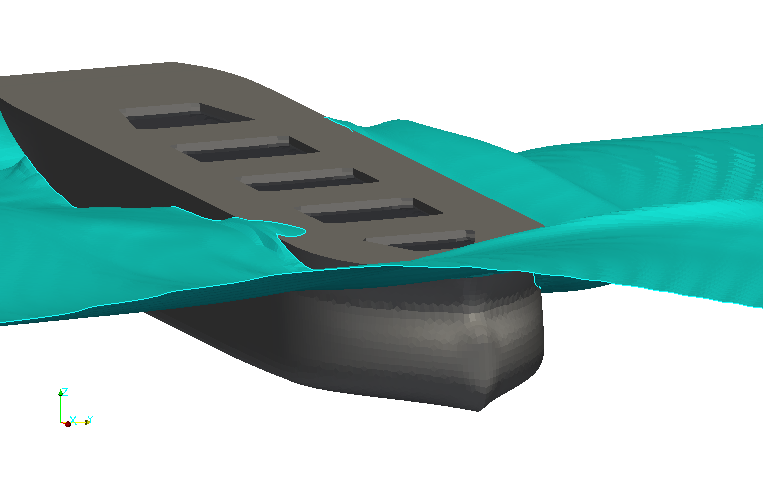}
    }
    \caption{Snap shots of bulk carrier in seaway during incident conditions as experienced by "Jian Fu Star".}
 \label{3D_vessel}
\end{figure}

The hydrodynamic response of a cargo vessel subjected to wave-induced motions was computationally simulated for a duration of $40.0$ seconds utilizing high-performance computing (HPC) infrastructure. This complex numerical simulation necessitated a substantial computational expenditure, consuming approximately $18\cdot 10^4$ Central Processing Unit (CPU) hours to complete the prescribed $40.0$ second time domain analysis.\\
In Fig. \ref{Holds3DViscos}, the viscosity of the cargo calculated using the Drucker-Prager yield criterion is visualized for regions where the soil mixture fraction is between $0.95$ and $1.0$. At time $t=3.0$\;s, the cargo loading configuration results in circular patterns of lower viscosity, indicative of the cargo's propensity to spread and flatten at the beginning of the simulation. As the simulation progresses to $t=18.5$\;s, the cargo has already undergone significant flattening. Concurrently, the vessel's pitch motion induces high accelerations in hold one. Consequently, the viscosity exhibits a gradient, with the lowest values observed in the forward hold and progressively increasing towards the aft holds. This general trend persists in subsequent time steps. At the time $t=35.0$\;s, a cargo sliding plane can be observed in hold four, and generally, a cargo shift can be observed at the top of the cargo piles. This shift stays prominent until time $t=39.5$\;s, where a slight shift to the front of the vessel can also be detected. The spatial and temporal variations in viscosity can be attributed to several factors: initial cargo distribution, gravitational effects, vessel motion, and interaction between cargo and the hold structures.\\
The three degrees of freedom (3DoF) motions exhibited by the vessel during the simulation are illustrated in Fig. \ref{3DMotion}. Two of the snapshots presented in Fig. \ref{Holds3DViscos} were strategically selected to depict the vessel's orientation at the extrema of its pitch angle oscillations ($t=18.5$\;s and $t=35.0$\;s). The observed processive rolling motion of the vessel suggests a high probability of cargo shift occurrence, potentially leading to capsizing. Upon closer examination of the uppermost cells in Fig. \ref{Holds3DViscos}, a subtle yet discernible cargo displacement towards the starboard side becomes apparent from approximately $15.5$\; seconds into the simulation onwards.\\
\begin{figure}[htbp]
\centering
  \subfigure[$15.50$\;s]{ 
\includegraphics[scale=.38]{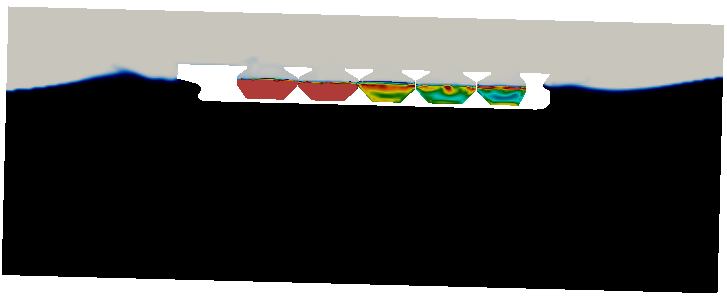}
   }
  \subfigure[$18.75$\;s]{ 
\includegraphics[scale=.38]{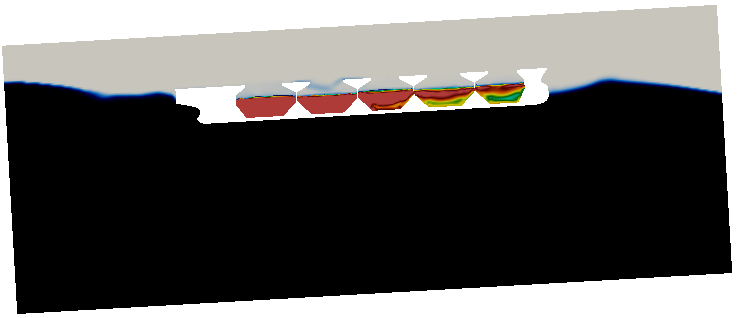}
   }
    \subfigure[$21.50$\;s]{ 
\includegraphics[scale=.38]{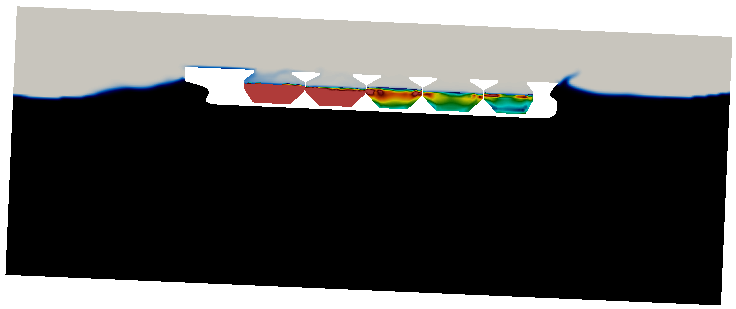}
   }
    \subfigure[$40.00$\;s]{ 
\includegraphics[scale=.40]{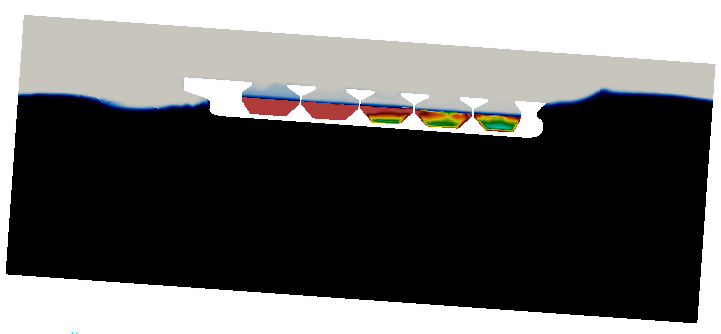}
   }
   \includegraphics[scale=.8]{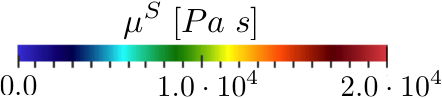}
    \caption{Field values of cargo viscosity in the $x_1$-$x_3$-plane at $x_2=0$\;m are depicted at four distinct time intervals for the 3D cargo vessel in wave reconstruction case. The black regions denote the water phase, while the grey regions indicate the air phase. The computational domain is dynamically rotating in accordance with the vessel's rigid body motions.}
 \label{MidschifsViskositaet}
\end{figure}
The viscosity distribution within the cargo holds is illustrated in Fig. \ref{MidschifsViskositaet} through a slice of the computational domain at $x_2=0$\;m. Also, the water and air phases are displayed, visually representing the vessel's position in the waves. The water phase is depicted in a spectrum from black to blue. At time points $t=18.75$\;s, $t=21.5$\;s, and $t=40.0$\;s, it is evident that the vessel is taking on water, as can also be observed in Fig. \ref{3D_vessel}. This observation aligns with the incident report of the "Jian Fu Star," although it should be noted that, in reality, cargo hold lids prevent water from sloshing into the holds. Due to the immiscible formulation of the cargo model, this water ingress does not affect the cargo properties. Nevertheless, at time $t=40.0$\;s, a second free surface inside the cargo holds can be observed, which will highly affect the vessel's stability.\\
\newpage
Figure \ref{MidschifsViskositaet} also reveals that the cargo exhibits softer behavior in the forward holds. This phenomenon is particularly pronounced when the vessel is pitching downward, as exemplified by the pitch angle maxima at $t=21.5$\;s, where the viscosity reaches its minimum at the bottom of hold one.\\
Lee's study \cite{Lee2017} on the "Alam Manis" vessel, which traversed a typhoon, provides a detailed account of cargo consistency in each hold. The findings of that study corroborate the results of the present feasibility study. Specifically, the highest accelerations were observed to act on the cargo in the forward holds, with the most significant cargo shifts occurring in hold one. Lee \cite{Lee2017} reported a gradual decrease in cargo shift angle from the forward to the aft holds. Furthermore, the IMO "Technical Working Group" \cite{TWG2013} confirms that the highest accelerations during bulk carrier voyages were recorded in the forward hold, which is consistent with the results presented in this study.\\
\begin{figure}[t]
\centering
  \subfigure[Hold 1 ($x_1=162$\;m)]{ 
\includegraphics[scale=.18]{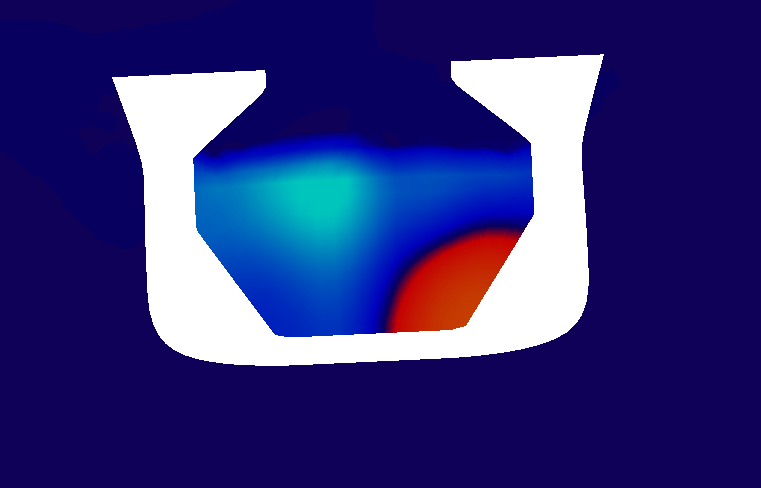}
   }
  \subfigure[Hold 2 ($x_1=134$\;m)]{ 
\includegraphics[scale=.18]{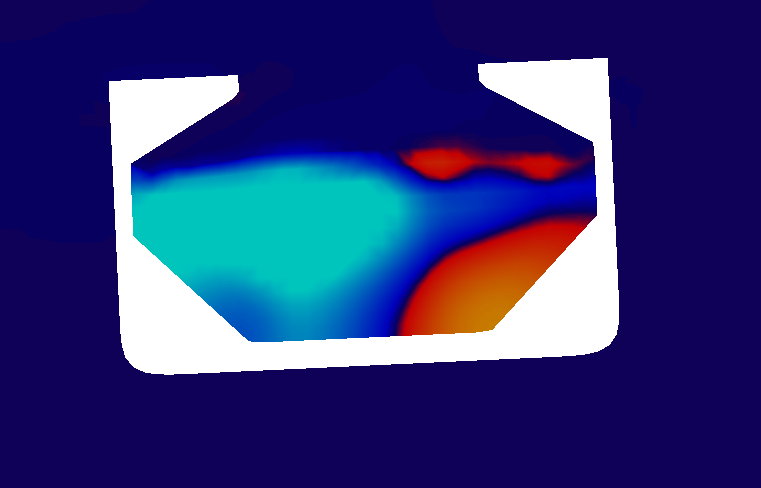}
   }
    \subfigure[Hold 3 ($x_1=104$\;m)]{ 
\includegraphics[scale=.18]{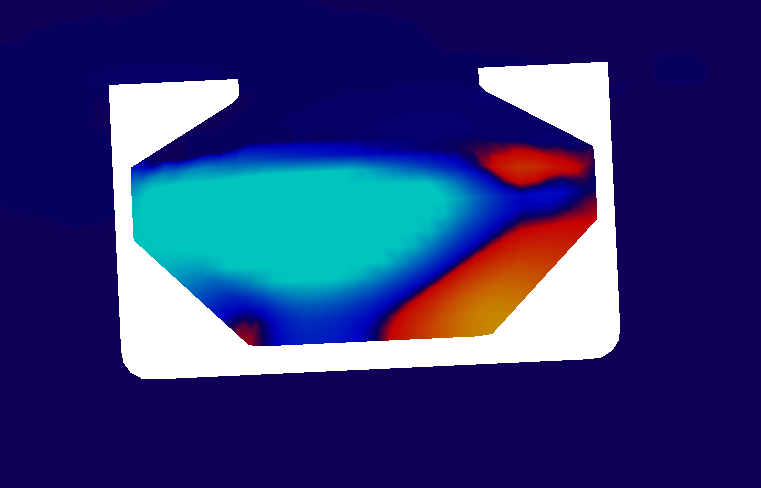}
   }
     \subfigure[Hold 4 ($x_1=74$\;m)]{ 
     \includegraphics[scale=.18]{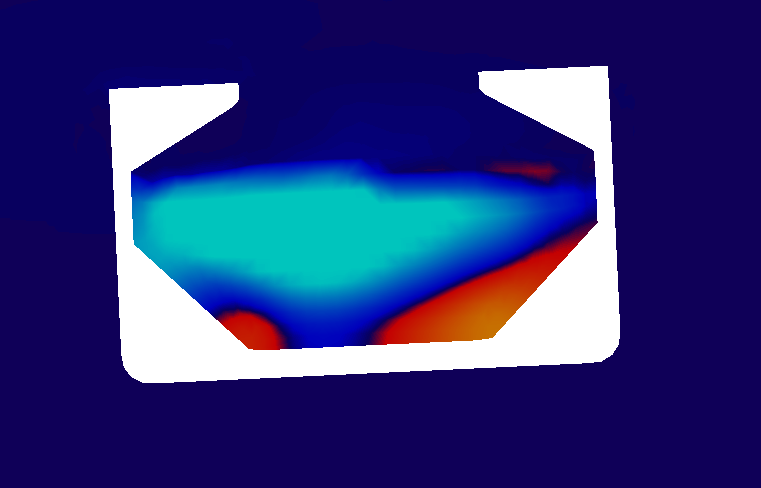}
   }
      \subfigure[Hold 5 ($x_1=40$\;m)]{ 
\includegraphics[scale=.18]{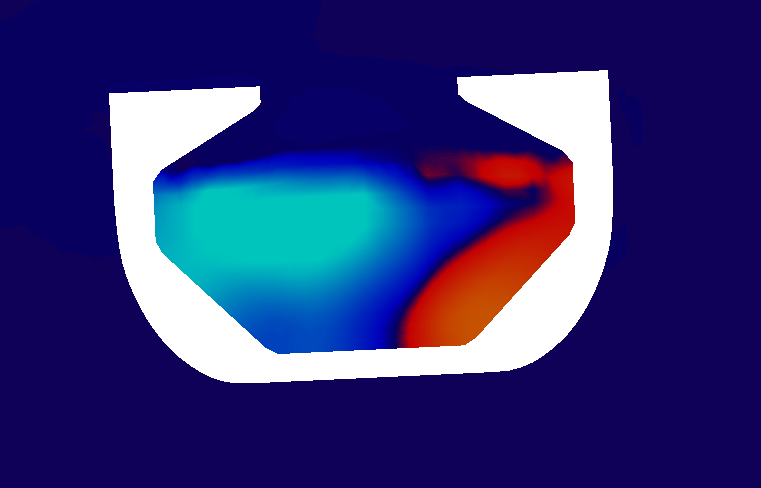}
   }
   \includegraphics[scale=.65]{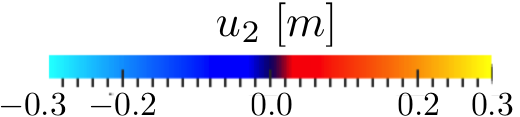}
    \caption{Cargo displacement in the $x_2$-direction at time $t=30.0$\;s displayed on $x_2$-$x_3$-planes through all five holds for the 3D cargo vessel in wave reconstruction case.}
 \label{DisplY_30s}
\end{figure}
\begin{figure}[htbp]
\centering
  \subfigure[$t=1.0$\;s]{ 
  \includegraphics[scale=.24]{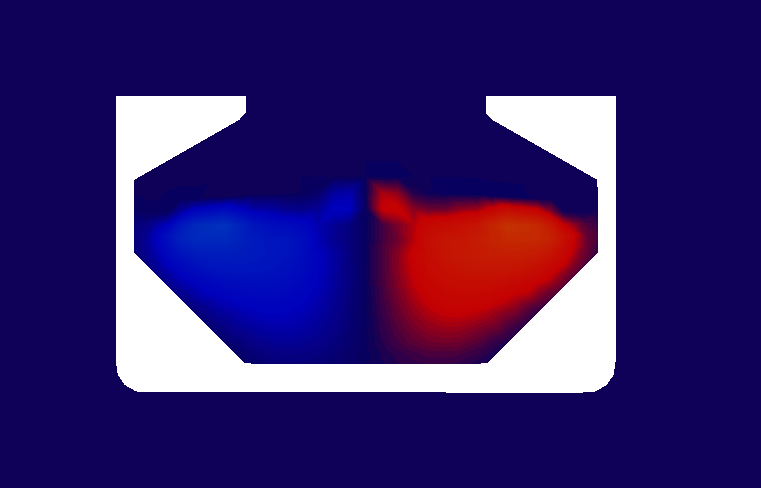}
    }
  \subfigure[$t=15.5$\;s]{ 
  \includegraphics[scale=.24]{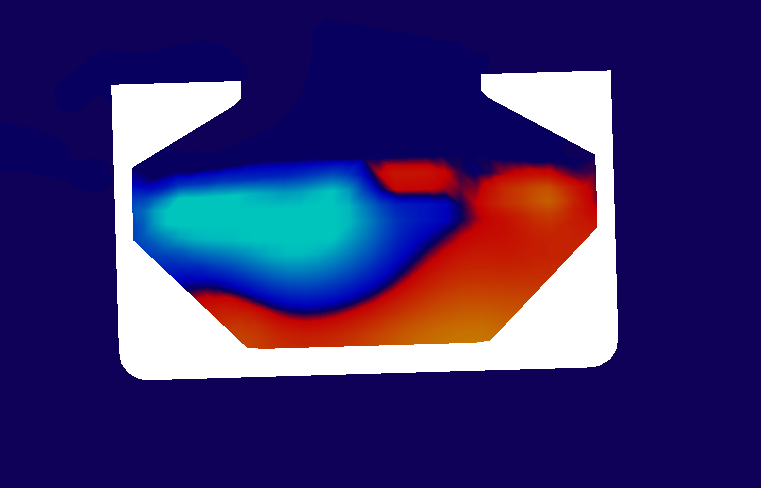}
   }
     \subfigure[$t=30.0$\;s]{ 
\includegraphics[scale=.24]{./Bilder/Hold4_disply_30s.png}
   }
      \subfigure[$t=40.0$\;s]{ 
\includegraphics[scale=.29]{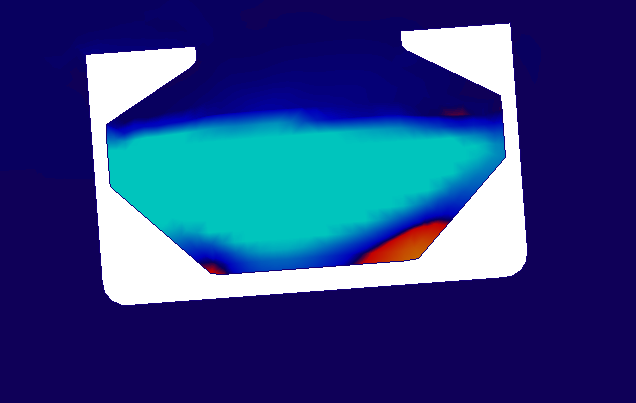}
   }
   \includegraphics[scale=.65]{./Bilder/displ_2_3Dvessel.png}
    \caption{Cargo displacement in the $x_2$-direction at four time steps displayed on $x_2$-$x_3$-planes in hold 4 for the 3D cargo vessel in wave reconstruction case.}
 \label{DisplY_74m}
\end{figure}
\begin{figure}[htbp]
\centering
  \subfigure[rear view]{ 
  \includegraphics[scale=.18]{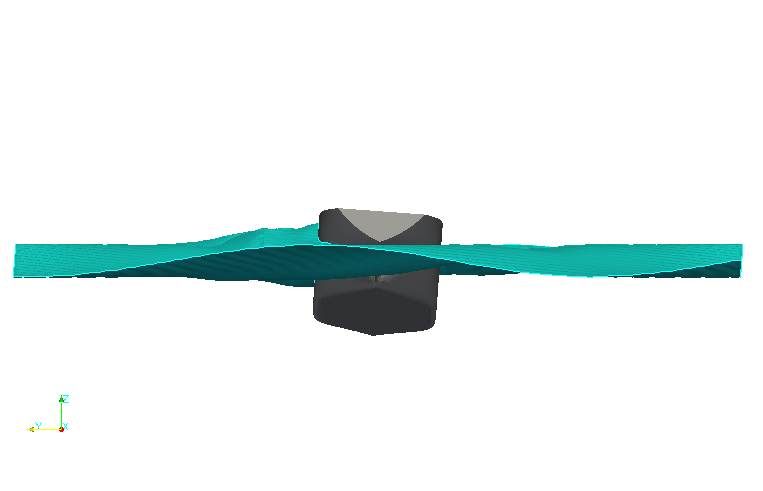}
    }
     \subfigure[side view]{ 
\includegraphics[scale=.18]{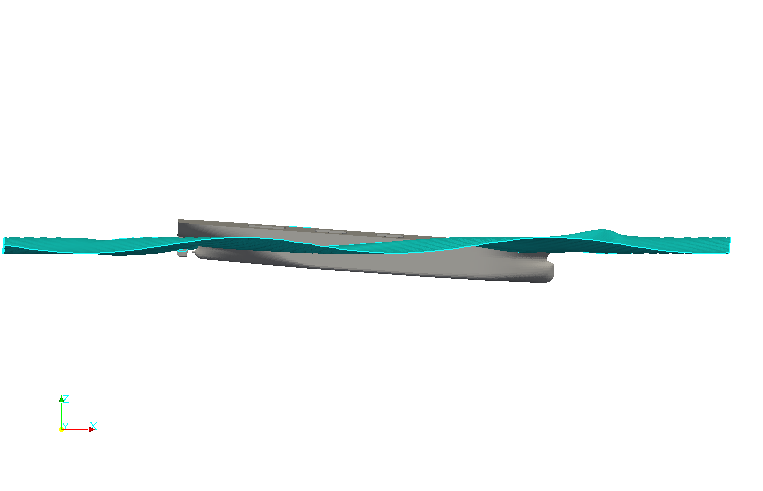}
   }
      \subfigure[front view]{ 
\includegraphics[scale=.18]{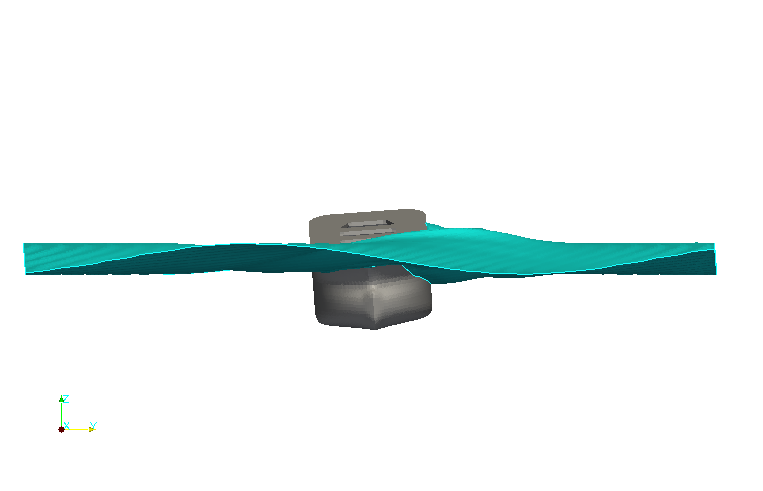}
   }
    \caption{Snap shots of the bulk carrier at time $t=40.0$\;s from the front, side and rear view.}
 \label{3D_40s}
\end{figure}
To elucidate the impact of the vessel's successive rolling motion on the cargo, as illustrated in Fig. \ref{3DMotion}, Fig. \ref{DisplY_74m} presents the second component of the Eulerian cargo displacement at four distinct time points across a plane intersecting hold four. At $t=15.5$\;s, $t=30.0$\;s and $t=40.0$\;s, a notable shift towards starboard (indicated by negative displacement) is observed. Over time, the line of zero displacement progressively migrates towards the vessel's port side (right side in the figures).\\
In Fig. \ref{DisplY_30s}, the Eulerian cargo displacement is visualized across planes intersecting each vessel's hold at time $t=30.0$\;s. Significantly, the most substantial displacement is manifested in holds two and three, specifically in the central regions of these cargo compartments. This phenomenon can be attributed to two primary factors: 
\begin{itemize}
\item{The geometric constraints imposed by the hold structures, which restrict the extent of displacement.}
\item{The reduced viscosity in the central areas of the holds, as evidenced in Figs. \ref{Holds3DViscos} and \ref{MidschifsViskositaet}.}
\end{itemize} 
The final position of the bulk carrier in the simulation is depicted in Fig. \ref{3D_40s} for three different views: front, rear, and side view. The immense pitch of the vessel, which could result in a sinking over the nose of the vessel, can be observed, indicating a possible slight cargo shift to the front of the vessel. Another mechanism that could be responsible for the reinforced pitch of the vessel is the water film on top of the cargo indicated in Fig. \ref{MidschifsViskositaet}.\\
Snap shots of the bulk carrier during the simulation are displayed in Fig. \ref{3D_vessel}, visualizing the 3D interaction of the hull and the water surface. Clearly, at $t=20.0$\;s water flows in the rear hold. Also, at $t=25.0$\;s and $t=39.5$\;s water can be observed on deck of the vessel.\\
The feasibility study demonstrates the efficacy of a monolithic coupled 3D model in replicating observed cargo behavior using a rigid perfectly-plastic material approach. This methodology accurately captures the mechanical behavior of materials that do not exhibit strain hardening or strain-rate sensitivity during yielding. However, the direct monolithic coupling of multi-physics processes with disparate time scales, encompassing cargo behavior on a vessel in waves, necessitates substantial computational resources. The computational demands stem from several factors:
\begin{itemize}
\item{Monolithic Coupling: The integration of multiple physical processes into a single, unified model.}
\item{Time Scale Disparities: The need to simultaneously resolve phenomena occurring at different temporal resolutions.}
\item{3D Modeling: The increased complexity associated with three-dimensional simulations.}
\end{itemize} 
Expanding the 3D approach to incorporate coupled seepage flow through rigid perfectly-plastic materials would further enhance the model's capabilities. This extension could account for: "wet base" and "dynamic separation" scenarios. However, such an expansion would significantly increase computational costs due to:
\begin{itemize}
\item{Doubled Governing Equations: The need to solve twice the number of equations governing the system.}
\item{Increased Complexity: The addition of fluid-structure interaction in porous media.}
\end{itemize}

\section{Conclusion}
\label{Concl}
The introduced monolithic numerical model for granular cargo transport on bulk carriers is validated, verified, and applied to a real-case scenario. This advanced numerical framework offers a robust and efficient method for simulating the complex interplay between vessel motion, wave dynamics, and granular cargo behavior. Its application to the "Jian Fu Star" incident demonstrates its potential for enhancing maritime safety through an improved understanding of bulk carrier stability and cargo shifting phenomena.\\
The validation and verification studies demonstrate the robustness and accuracy of the three-phase flow method and the granular material model. For the three-phase flow, the numerical results showed strong agreement with experimental data regarding front position and overall flow behavior. While some discrepancies were observed in breaking behavior and velocity profiles, likely due to differences in initial conditions, the model successfully captured the key dynamics of the dam break experiment.\\
When compared to pFEM results, the granular material model exhibited excellent concordance in flow patterns and final deposition shapes. The slight differences observed in wall adhesion highlight the different treatment of boundary interactions between the two methods.\\
The feasibility study shows a cargo shift occurring in the assumed weather and wave conditions while using material properties and vessel and cargo dimensions as given in the incident report. Since a cargo shift is observed with a rigid, perfectly-plastic model, the question arises if cargo liquefaction was the underlying reason for the "Jian Fu Star" sinking or if a different phenomenon is responsible for the accident. These observations underscore the complex interplay between vessel motion, cargo characteristics, and hold geometry in determining the spatial distribution and magnitude of cargo displacement during maritime transport.\\
While the monolithic coupled 3D approach with rigid perfectly-plastic materials shows promise in reproducing cargo behavior, its computational intensity presents challenges for practical implementation. While theoretically beneficial, the proposed extension to include seepage flow would exacerbate these computational demands. Future research should focus on balancing model fidelity with computational efficiency to make such comprehensive simulations more feasible for real-world applications.

\section*{Acknowledgements}
The authors acknowledge the support within the research project ``LiquefAction - Cargo Liquefaction in Ship Design and Operation'' (Grant No. FKZ 03SX363A ). 
Selected computations were performed with resources provided by the North-German Super-computing Alliance (HLRN). 

\bibliography{BibPerfPlastic}
\bibliographystyle{plain} 

\section{Appendix}
\subsection{Vessel dimensions of "Jian Fu Star" model}

\end{document}